%% file: main_document_no_red.tex
\documentclass[aps,pre,reprint,twocolumn,showpacs,superscriptaddress,amsmath,amssymb]{revtex4-1}
\usepackage{graphicx}  % needed for figures5
\usepackage{dcolumn}   % needed for some tables
\usepackage{bm}        % for math
\usepackage{hyperref}
\usepackage{verbatim}  % for comment
\usepackage{natbib}
\usepackage{epsfig}
\usepackage{epstopdf}
\usepackage{booktabs}  % for multicolumn 
\usepackage{color}
\usepackage[toc,page]{appendix}
\usepackage{placeins}
\usepackage{float}
\usepackage{multirow}
\usepackage[normalem]{ulem}
\usepackage{amsmath}
\usepackage{tabu}
\usepackage[utf8]{inputenc}
\usepackage{rotating}
\usepackage{multirow}
\usepackage[export]{adjustbox}

\begin{document}

\title{Nucleation on a sphere: the roles of curvature, confinement and ensemble}
\date{\today}

\author{Jack O. Law}
\email{jack.o.law@durham.ac.uk}
\affiliation{Department of Physics, Durham University, South Road, DH1 3LE}
\author{Alex G. Wong}
\affiliation{Department of Chemistry, Durham University, South Road, DH1 3LE}
\author{Halim Kusumaatmaja}
\email{halim.kusumaatmaja@durham.ac.uk}
\affiliation{Department of Physics, Durham University, South Road, DH1 3LE}
\author{Mark A. Miller}
\email{m.a.miller@durham.ac.uk}
\affiliation{Department of Chemistry, Durham University, South Road, DH1 3LE}

\begin{abstract}
By combining Monte Carlo simulations and analytical models, we demonstrate and explain how the gas-to-liquid phase transition of colloidal systems confined to a spherical surface depends on the curvature and size of the surface, and on the choice of thermodynamic ensemble. We find that the geometry of the surface affects the shape of the free energy profile and the size of the critical nucleus by altering the perimeter--area ratio of isotropic clusters. Confinement to a smaller spherical surface results in both a lower nucleation barrier and a smaller critical nucleus size. Furthermore, the liquid domain does not grow indefinitely on a sphere. Saturation of the liquid density in the grand canonical ensemble and the depletion of the gas phase in the canonical ensemble lead to a minimum in the free energy profile, with a sharp increase in free energy for additional growth beyond this minimum. 
\end{abstract}


\maketitle

\section{\label{sec:introduction} Introduction}
Two-dimensional systems are relevant to a wide range of problems in nature and engineering, and phase transitions in such systems are strongly affected by any curvature of the surface to which they are confined. During nucleation and growth, curvature can not only influence the size and shape of the critical nucleus, but also fundamentally alter the nucleation pathway~\cite{Gomez2015A,Marenduzzo2012A,Meng2014A,Yao2017A}. These effects have a profound impact on a number of self-assembling systems in nature, such as the development of the HIV-1 virus capsid~\cite{Paquay2016A,Li2000A,Perotti2016A,Zlotnik2005A} and the assembly of clathrin-coated pits on cell membranes~\cite{Vliegenthart2011A,Tamotsu2004A}. During spinodal decomposition, confinement to a curved surface can introduce new metastable states~\cite{Bott2016A,Paillusson2016A} and even suppress typical spinodal decomposition in favour of Ostwald ripening~\cite{Bott2016A}. This may influence the formation of lipid rafts in cell membranes~\cite{Bott2016A}, which are thought to be responsible for inter-cell interaction and may drive endocytosis~\cite{Nicolas2009A,Simons2010A,Goryachev2011A}. In technological applications, curved colloidal systems are now exploited for the development of novel smart materials, with applications including encapsulation~\cite{Joshi2016A,Bausch2003A,Dinsmore2002A}, soft lithography~\cite{Park1997A,Kramer2005A}, defect functionalisation~\cite{DeVries2007A} and the creation of artificial cells~\cite{Mohrdieck2007A,Limozin2005A}. Recent research into these systems is further buoyed by the success of novel techniques for fine control over colloidal loading and interactions~\cite{Joshi2016A,Dinsmore2002A,Meng2014A,Fantoni2012}, as well as the shape of the confining surface~\cite{Kusumaatmaja2013A,Irvine2010A,Pairam2014A}.

When investigating any phase transition on curved two-dimensional systems, there are several important effects which are not always easy to uncouple. Firstly, the domain shape with the smallest boundary length has a perimeter-area relation that is different from that of a circular disk for a flat surface.
Secondly, it is possible for a curved surface (such as the surface of a sphere) to have a finite area without being bounded by an edge.  In contrast, a planar surface must either be notionally infinite or delimited by an edge that particles cannot cross.  The edgeless yet finite space defined by some curved surfaces is a key feature~\cite{Rao1979A,Bott2016A}. When dealing with phase transitions in computer simulations, the usual practice is to minimise finite-size effects by extrapolating to the thermodynamic limit \cite{Caillol1998A,Wilding1995A}, but this is no longer appropriate in an intrinsically finite space. Thirdly, the equivalence between the canonical and grand canonical ensembles is broken. Thus, it is important to distinguish between open and closed experimental systems. 

To provide insights into these effects, and in particular to isolate their individual significance, we investigate what is perhaps the simplest phase transition in a curved two-dimensional system: the gas-to-liquid nucleation of colloidal particles confined to a sphere. In this work we use a combination of particle-based Monte Carlo simulations and analytical models. In agreement with previous continuum models~\cite{Gomez2015A,Marenduzzo2012A,Koehler2016A}, we show the surface geometry primarily affects the nucleation barrier and critical nucleus by altering the relationship between its perimeter and area. Our Monte Carlo simulation results further confirm G{\'o}mez's prediction~\cite{Gomez2015A} for the dependency of the size of the critical nucleus on the curvature except at low supersaturations, where the line tension becomes more sensitive to curvature.

In our Monte Carlo simulations we also observe that the liquid domain does not grow indefinitely as there is a minimum in the free energy profile. In the grand canonical ensemble this is caused by the saturation of the liquid density, while in the canonical ensemble it is due to the depletion of the gas phase---an effect that has been exploited in the liquid--solid transition in three-dimensional systems to control polymorph crystallisation~\cite{Chen2011B,Nicholson2011}. The aforementioned continuum models ignore these considerations. By assuming the liquid phase to be a van der Waals-like fluid and the gas phase to be an ideal gas, we can develop an analytical model that captures the key features of the Monte Carlo simulation results in the grand canonical ensemble. Finally, we shed some light onto the difference between simulating in the canonical and grand canonical ensembles. In particular we observe that nucleation is absent for a Lennard-Jones system in the canonical ensemble. However, we are able to characterise the free energy of a liquid cluster at around its equilibrium size, and explain the shape of the curve by extending a free energy model for nucleation in finite systems developed by Rao and Berne~\cite{Rao1979A} and by Reguera {\it et al.}~\cite{Reguera2003}.

This paper is organised as follows. In Sec.~\ref{sec:methods} we describe the model system to be simulated and the methods employed to generate and analyse our results. In Sec.~\ref{sec:GrandCanonical} we present and discuss our results for the grand canonical ensemble. This is followed by the results for the canonical ensemble is Sec.~\ref{sec:canonical}. Finally, in Sec.~\ref{sec:conclusions} we draw conclusions and consider directions for future work.

\section{\label{sec:methods} Model and Methods}

In order to understand the gas--liquid phase transition on the surface of a sphere we will study a model system of Lennard-Jones particles confined to a spherical surface. The following tools are required to analyse this model: (i) a Monte Carlo algorithm that correctly and efficiently samples states of the spherical system; (ii) the multiple histogram reweighting technique, which allows us to find the gas--liquid coexistence curve in the grand canonical ensemble; (iii) a reaction coordinate, to measure the free energy profile of the nucleation process; and (iv) the umbrella sampling technique, to aid the sampling of states along the nucleation path.

\subsection{\label{sub:modelSystem} Model system}

Our model system consists of soft spherical particles whose centres are confined to the surface of a sphere of radius $R$. The particles interact isotropically {\textit{via}} the truncated, shifted and smoothed Lennard-Jones potential,
%
%\begin{widetext}
\begin{displaymath}
	U(r) = \left[U_{\text{LJ}}(r)-U_{\text{LJ}}(r_{\text{c}})-(r-r_{\text{c}})\frac{dU_{\text{LJ}}}{dr}\bigg|_{r_{\text{c}}}\right]H(r_{\text{c}}-r),
\end{displaymath}
%\end{widetext}
%
where $r$ is the separation of the two particles, measured in three-dimensional space, rather than along the geodesic; $r_{\text{c}}$ is the cut-off distance, at which the potential is truncated; and $H(x)$ is the Heaviside function~\cite{Paquay2016A}. $U_{\text{LJ}}(r)$ is the Lennard-Jones potential, given by
\begin{displaymath}
	U_{\text{LJ}}(r) = \varepsilon\left[\left(\frac{r_{\text{m}}}{r}\right)^{12}-2\left(\frac{r_{\text{m}}}{r}\right)^6\right],
\end{displaymath}
where $\varepsilon$ is the depth of the potential well and $r_{\text{m}}$ is the separation at which the potential reaches its minimum value~\cite{Jones1924A}. We choose to fix $r_{\text{c}}/r_{\text{m}}=2.23$ throughout this work, in order to match the potential chosen by other authors~\cite{Paquay2016A,Gribova2011A,tenWolde1998A,Rao1979A,Vest2014A}.
We note that previous work on planar two-dimensional Lennard-Jones systems has shown that the phase diagram is significantly affected by the choice of potential truncation \cite{Smit91a}.

For the remainder of this work, all quantities are reported in reduced units where energy is measured in units of $\varepsilon$, distance in units of $r_{\text{m}}$ and temperature in units of $\varepsilon/k_{\rm B}$, where $k_{\rm B}$ is Boltzmann's constant.

\subsection{\label{sub:monteCarlo}Monte Carlo simulation}

\begin{figure}
\includegraphics[width=0.48\textwidth]{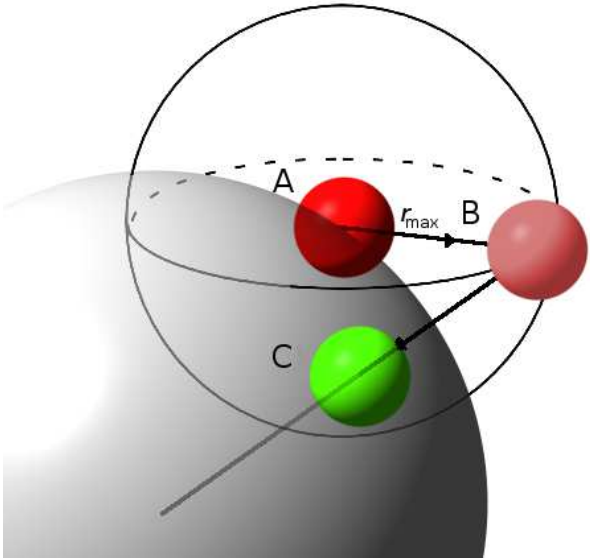}
\caption{\label{fig:move} Schematic of trial Monte Carlo moves. The particle is initially at position A, on the spherical surface shown in grey. It is moved to a random point B on the surface of a small sphere of radius $r_{\text{max}}$, shown here as a wire-frame. It is then projected radially back onto the confining surface. Therefore, its trial position for the Monte Carlo move is at position C.}
\end{figure}

We use the Metropolis Monte Carlo algorithm to perform simulations in both the canonical ($NVT$) and grand canonical (${\mu}VT$) ensembles~\cite{Frenkle2002A,Metropolis1953A}. In order to sample the configuration space of particles confined to a sphere correctly and efficiently the following scheme, depicted in Fig.~\ref{fig:move}, is adopted~\cite{Brannon2002A}. Firstly, a randomly selected particle is moved onto the surface of a small sphere of radius $r_{\text{max}}$ centred on its old position, where $r_{\text{max}}$ is the maximum move length. The particle's new position on the small sphere is generated at random using the Marsaglia method~\cite{Marsaglia1972a}. Then the particle is projected radially back onto the confining sphere. The value of $r_{\text{max}}$ is adjusted throughout the equilibration stage of the simulation until the probability of a move being accepted is approximately 0.25, which ensures that the sampling of phase space is good~\cite{Mountain1994A}. This method both correctly samples the sphere~\cite{Brannon2002A}, and produces a computational performance improvement of about a factor of two over making moves uniformly distributed in the polar angle $\phi$ and in $\cos{(\theta)}$, where $\theta$ is the azimuthal polar coordinate.

To improve computational efficiency further, replica exchange (also known as parallel tempering) is employed~\cite{Swendsen1986A,Yan1999A}. During simulations aimed at locating the coexistence curve, exchange moves are made between temperatures. When using umbrella sampling, exchange moves are made between sizes of the target nucleus (see Sec.~\ref{sub:umbrella}). To improve the efficiency of the canonical simulations, collective moves are used. These are based on the scheme of Troisi {\it et al.}~\cite{Troisi2005A}, modified to include the fluctuating pseudo-temperature described by Whitelam and Geissler \cite{Whitelam2007A}.

\subsection{\label{sub:multipleHistogram} Multiple histogram reweighting}

Grand canonical Monte Carlo simulations can be used to trace the gas--liquid coexistence curve. To explain how this is done, we first note that in the grand canonical ensemble, the system has a fixed temperature $T$, and activity $z$, which is defined as
\begin{equation}
	z(\mu)=\frac{A_{0}}{\Lambda^2}\exp\left(\frac{\mu}{k_{\rm B}T}\right),
\label{zdef}
\end{equation} 
where $\mu$ is the chemical potential of the particle reservoir, $A_{0}$ is the area of the confining sphere and $\Lambda$ is the thermal de Broglie wavelength. For a given $(z,T)$ pair, the probability density of observing the system containing $N$ particles and having energy $E$ is
\begin{equation}
	h\left(N,E|z,T\right) = \frac{\Theta\left(N,E\right)}{\Xi\left(z,T\right)} z^N \exp\left(-E/k_{\text{B}}T\right),
\label{probN}
\end{equation}
where $\Theta\left(N,E\right)$ is the density of states and $\Xi\left(z,T\right)$ is the grand canonical partition function. We measure the density and energy histogram $h\left(N,E | z,T\right)$ for a number of temperature--activity pairs close to coexistence. These histograms can be collectively fitted to Eq.~(\ref{probN}) to generate a self-consistent approximation of $\Theta\left(N,E\right)$. This density of states can then be used to locate coexistence histograms, where the weights of the gas and liquid peaks are equal, and thereby trace the coexistence curve~\cite{Wilding2001A,Weerasinghe1993A,Miller2004A}.

\subsection{\label{sub:reactCoord}The reaction coordinate}
In order to measure the free energy barrier to nucleation, a reaction coordinate must be chosen. Firstly, we will describe our choice of reaction coordinate for simulations in the context of the grand canonical ensemble, then discuss how it can be extended to the canonical ensemble. In the grand canonical ensemble, it can be assumed that the nucleation pathway will consist of the growth of liquid clusters against a background of gas, which maintains a constant density thanks to the availability of the particle bath. Therefore, the free energy change associated with the addition of a liquid cluster of size $N$ is simply the free energy difference between a cluster of size $N$ and area $A$ and a gas occupying area $A$. Each cluster in the system can be considered independently. The reaction coordinate is defined for each cluster and is simply the number of particles in a cluster~\cite{Stillinger1963A,Santra2008A,tenWolde1998A}.

We use a definition based on pairwise separations to define a liquid cluster \cite{Stillinger1963A}: any pair of particles that can be linked by a chain of particles in which each link is shorter than a cut-off distance $r_{\text{c}}$ are in the same cluster, while all other pairs are not. During the simulation, the cluster size distribution is measured every 2000 Monte Carlo steps. The free energy of a cluster containing $N$ particles can be calculated using 
\begin{equation}
	F(N) = -k_{\text{B}}T\ln{\frac{M_N}{\sum\limits_{s=0}^Q N_{\text {Total}}\left(s\right)}}, \label{grandfree}
\end{equation}
where $M_N$ is the total number of times a cluster of size $N$ is observed, $N_{\text {Total}}(s)$ is the total number of particles present on the sphere at measurement $s$, and $Q$ is the total number of measurements made during the simulation.

The methods described above can also be applied to the canonical ensemble under certain conditions. In the canonical ensemble, as the nucleus grows, the gas phase is depleted. Therefore, if there is more than one nucleus in the system, the nuclei's free energies are not independent and they should not be considered separately. This means that, in the canonical ensemble, this definition of the free energy of a cluster can only be applied to situations in which it is very unlikely that more than one nucleus will be found in the system.

\subsection{\label{sub:umbrella} Umbrella sampling}

When the barrier is low, the method of calculating the free energy profile described in Sec.~\ref{sub:reactCoord} is sufficient. However, when the barrier is high, clusters at the top of the barrier are rarely seen and so this region is poorly sampled. In these cases, umbrella sampling is needed~\cite{tenWolde1998A,Filion2010A,Zhong-can1987A}. Here, a fictitious biasing potential, which favours systems containing otherwise unlikely clusters, is added to the normal interaction potential. This bias can then be factorised out of the resulting histograms. We chose the biasing potential 
\begin{displaymath}
U_{\text{bias}} = \frac{1}{2}\lambda\left(N - N_{\text{target}}\right)^2,
\end{displaymath}
where $N$ is the number of particles in the largest cluster, $N_{\text{target}}$ is the cluster size which we would like to bias the system towards, and $\lambda$ determines the strength of the biasing potential. For this work we found that $\lambda = 0.024$ produced good results. 

In the canonical ensemble, the utility of umbrella sampling is somewhat limited. In order for the equilibrium state to contain a significant area of both phases, the system must be prepared as a deeply supersaturated gas, well into the spinodal region. One must be careful when applying umbrella sampling to such systems. Consider a system containing a small cluster, held within a narrow size range by umbrella sampling. The remaining particles are assumed to form a gas around the cluster. However, in the canonical ensemble, it is possible that this gas may still be supersaturated if the number of particles in the system is high enough. In this case the gas will collapse into additional liquid clusters. Therefore, the pathway mapped out by successive umbrella sampling runs will not represent the real transition pathway, which is simply spinodal decomposition. However, when the target nucleus is larger, so that the remaining gas is dilute enough that it does not spontaneously condense, umbrella sampling can be safely used. Therefore, one can map the free energy of liquid clusters around the equilibrium cluster size.

\section{\label{sec:GrandCanonical} The Grand Canonical Ensemble}

\subsection{\label{sub:phaseGC}Phase diagram}
\begin{figure}
\includegraphics[width=0.48\textwidth]{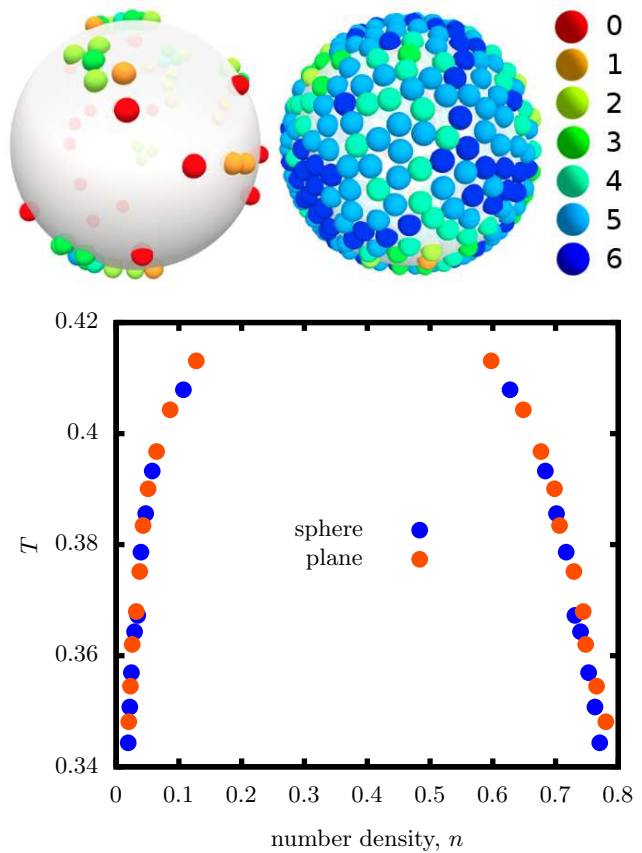}
\input{phase_diag}
\caption{\label{fig:coex}
Top: Snapshots of simulations on a sphere of radius $R=7$, in the gas and liquid phases from grand canonical simulations. Particles are coloured by their coordination number as indicated in the legend. Bottom: The coexistence curve of truncated, shifted and smoothed Lennard-Jones particles confined to the sphere (blue) and to a periodic square plane (orange) with equal area. It can be seen that the geometry of the surface does not strongly affect the gas--liquid coexistence curve in this system.}
\end{figure}

Nucleation is likely to be seen in conditions close to coexistence, so knowledge of the phase diagram is necessary for measurements of the nucleation barrier to be made. We employed the multiple histogram reweighting technique to measure the gas--liquid coexistence curve of our model in the grand canonical ensemble. We chose to do this for a sphere of radius $R=7$, which is small enough that the effects of curvature may be observed but large enough to hold sufficient particles for our measurements. We compare this phase diagram to that on a periodic square plane of the same area. It can be seen in Fig.~\ref{fig:coex} that these two phase diagrams are very similar. There are two reasons for this. The first is that both the gas and liquid phases are amorphous so, unlike for elastic structures such as crystals, little structural frustration is introduced by the curvature~\cite{Bausch2003A,Meng2014A,Koehler2016A,Azadi2014A,Lipowski2005A,Schneider2005A}. The second is that, for the overwhelming majority of the sampled states, the sphere is covered entirely in either liquid or gas, so there are no interfaces. Therefore, the overall geometry of the structure does not affect the phase behaviour (as it would if there were a phase boundary in the system, see Sec.~\ref{sub:nucGC}). For radii other than $7$ the same result was observed, except on spheres so small that they could only hold a few tens of particles. 

\subsection{\label{sub:nucGC}Nucleation}

\begin{figure}
	\input{ld_compare}
	\caption{\label{fig:compGC} An example nucleation curve taken from grand canonical simulations on a sphere of $R=7$ at $z = 20.90$. Fits to the CNT model assuming planar geometry, the G{\'o}mez modification to the line tension and the full spherical model are also shown.}
\end{figure}
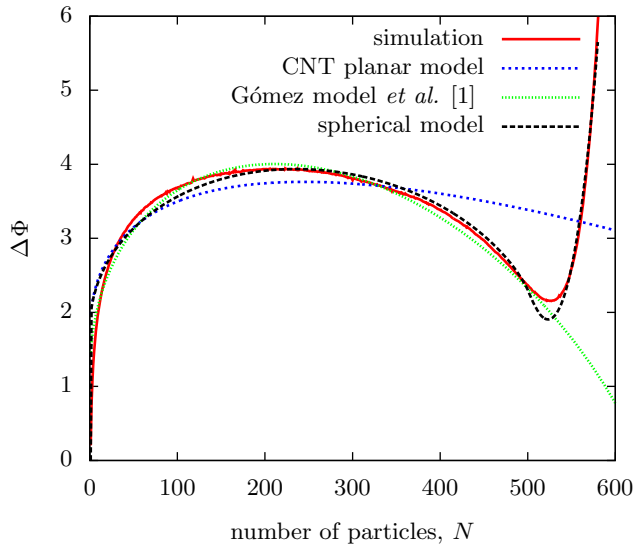

We measure the free energy barrier to the nucleation of the liquid phase on a sphere of radius $R=7$ with umbrella sampling. Fig.~\ref{fig:compGC} shows the resulting free energy as a function of the number of particles in the largest cluster, as defined in Sec.~\ref{sub:reactCoord}. The first feature of the free energy curve is a barrier, caused by the competition between the bulk free energy of the liquid phase and the line tension between the liquid and gas phases. The second feature is the minimum that is seen on the far side of the nucleation barrier. This is caused by the cluster completely covering the sphere. Adding further particles to the surface increases the density of the liquid phase above its equilibrium value, which has a high free energy cost.

In order to develop a model that explains these features, we start with the result from classical nucleation theory (CNT). In the grand canonical ensemble, the thermodynamic potential (free energy) is the grand potential $\Phi=F-\mu N$, where $F$ is the Helmholtz free energy.  The change in grand potential for creation of a nucleus is
\begin{equation}
\Delta \Phi = \gamma P + A_{\rm L}\left(P\right)\Delta \phi, \label{equ:CNT} 
\end{equation}
where $\gamma$ is the line tension, $\Delta\phi = \phi_{\text{L}} - \phi_{\text{G}}$ is the bulk free energy of the nucleating phase relative to the parent phase per unit area, $P$ is the perimeter of the nucleus and $A_{\rm L}\left(P\right)$ is the area of the nucleus as a function of its perimeter. Fig.~\ref{fig:compGC} includes an attempted fit of Eq.~(\ref{equ:CNT}) to the simulated results, assuming the usual planar relation $A_{\rm L}\left(P\right) = \pi{P^2}/4$. It can be seen that Eq.~(\ref{equ:CNT}) not only fails to capture the increase in the free energy at large cluster sizes, it also cannot fit the shape of the initial barrier.

Recently, G{\'o}mez {\it et al.}~developed a continuum Landau theory for phase nucleation on curved surfaces~\cite{Gomez2015A}, which predicts that, in the absence of elasticity or finite thickness, curvature affects the free energy barrier to nucleation by altering the relationship between the perimeter of the nucleus and its area. On a plane, assuming isotropic growth, $A\left(P\right) \propto P^2$. However, on surfaces with positive (negative) Gaussian curvature, the area grows faster (slower) as a function of $P$ than on a plane. This will naturally change the shape of the free energy profile described in Eq.~(\ref{equ:CNT}). On the sphere, the perimeter of a spherical cap of area $A_{\rm L}$ is given by
\begin{equation}
	P(A_{\text{L}}) = \sqrt{4\pi\left(A_{\text{L}} - \frac{A_{\text{L}}^2}{A_{\text{0}}}\right)},  \label{linetension}
\end{equation}
where $A_{\text{0}}$ is the total surface area of the sphere. As shown in Fig.~\ref{fig:compGC}, this modification is sufficient to capture the shape of the barrier. 

In order to model the increase in free energy caused by the spherical confinement, a density-dependent free energy is needed. This model is built on the assumptions (i) that the free energy can be written as the sum of a bulk term and a term describing line tension; (ii) that the liquid phase can be modelled as a van der Waals-like fluid; (iii) that the gas phase can be modelled as an ideal gas; (iv) that the density of the gas phase is constant (i.e., the gas is in chemical equilibrium with the particle bath throughout the nucleation process); and (v) that the line tension between the gas and liquid phases is constant.

Under these assumptions, the free energy of a liquid cluster is
\begin{equation}
	\Phi_{\text{L}} = -Nk_{\text{B}}T\left[1+\ln\left(\frac{A_{\text{L}}-Nb}{N\Lambda^2}\right)\right]-\frac{aN^2}{A_{\text{L}}} -\mu_{\rm L} N,
\label{PhiL}
\end{equation}
where $N$ is the number of particles in the liquid cluster, $A_{\text{L}}$ is the area of the liquid cluster, and $a$ and $b$ are the van der Waals parameters, representing the attraction between two particles, and the area occupied by a single gas particle respectively~\cite{Johnston2014A}. In Eq.~(\ref{PhiL}), the chemical potential of the liquid droplet is denoted $\mu_{\rm L}$ to allow for the fact that the droplet is not, in principle, in equilibrium with the particle bath.  The corresponding activity [see Eq.~(\ref{zdef})] is $z_{\rm L}=A_0 \Lambda^{-2} \exp(\mu_{\rm L}/k_{\rm B}T)$, so that $\Phi_{\rm L}$ may also be written as
\begin{displaymath}
	\Phi_{\text{L}} = -Nk_{\text{B}}T\left[1+\ln\left(\frac{A_{\text{L}}-Nb}{N{A_0}}\right)\right]-\frac{aN^2}{A_{\text{L}}} - Nk_{\rm B}T\ln(z_{\rm L}).
\end{displaymath}
It is the loss of area available to particles represented by the first logarithmic term in the expression for $\Phi_{\rm L}$ that will ultimately lead to the sharp increase in free energy of large liquid clusters.

As the liquid phase grows, it replaces an equal area of ideal gas with free energy
\begin{displaymath} 
	\Phi_{\text{G}} = -p A_{\rm L} = -n_{\rm G} A_{\rm L} k_{\rm B}T = -\frac{z}{A_0} A_{\rm L} k_{\rm B}T,
\end{displaymath}
where $n_{\rm G}$ is the number density of the gas phase.
As with CNT, the free energy cost of forming the interface between the gas and liquid phases is 
\begin{displaymath}
	\Phi_{\text{int}} = \gamma P(A_{\text{L}}).
\end{displaymath}

The total free energy of formation of a liquid cluster containing $N$ particles of size $A_{\text{L}}$ is

\begin{displaymath}
	\Delta \Phi = \Phi_{\text{L}} + \Phi_{\text{int}} - \Phi_{\text{G}} + d,
\end{displaymath}
where $d$ is a constant containing the translational entropy of the liquid cluster as a whole~\cite{Seunghwa2010A,Ford1997A,Girshick1990A}. The full form of the equation is then
\begin{widetext}
% \begin{equation}
% 	\Delta {\color{green}\Phi} = -Nk_{\text{B}}T\left[1+\ln\left(\frac{A_{\text{L}}-Nb}{N{\color{green}{\Lambda^2}}}\right)\right]-\frac{aN^2}{A_{\text{L}}} + \gamma\sqrt{4\pi}\sqrt{A_{\text{L}} - \frac{A_{\text{L}}^2}{A_{\text{0}}}} {\color{green} + (\mu n_{\rm G} - \mu n_{\rm L} - f_{\rm G})} A_{\text{L}} + d. \label{equ:grandFull} 
% \end{equation}
\begin{equation}
\begin{split}
\Delta \Phi = -N k_{\rm B} T \left[ 1 + \ln \left(\frac{A_{\rm L} - N b}{N A_0}\right)\right] - \frac{a N^2}{A_{\rm L}}
 - N k_{\rm B} T \ln\left(z_{\rm L}\right)
+ \gamma \sqrt{4 \pi \left(A_{\rm L}-\frac{A_{\rm L}^2}{A_0}\right)}
+ z k_{\rm B} T\frac{A_{\rm L}}{A_0} + d. \label{equ:grandFull}
\end{split}
\end{equation}
\end{widetext}

Eq.~(\ref{equ:grandFull}) has two mutually dependent variables: the cluster area $A_{\text{L}}$ and the number of particles it contains $N$. In order to fit the function to a simulated free energy profile measured using Eq.~(\ref{grandfree}), a function $A_{\text{L}}(N)$ is required. Rather than making a prediction for this function, we instead assume that for any $N$, the nucleus adopts the area that minimises its free energy, calculated by minimising Eq.~(\ref{equ:grandFull}) for $A_{\text{L}}(N)$. We find that the optimal $A_{\rm L}$ scales linearly with $N$, consistent with a constant liquid density $n_{\rm L}$, until the liquid covers the sphere.  $A_{\rm L}(N)$ then levels off at the total area $A_0$ of the sphere. 

Having optimised $A_{\rm L}$, the model has five fitting parameters: $a$, $b$, $d$, $z_{\rm L}$ and $\gamma$.  Although the parameters $a$ and $b$ are related via the second virial coefficient $B_2$ by $a/k_{\rm B}T=b-B_2$ in the exact van der Waals model, we find it necessary to treat $a$ and $b$ independently to capture the behaviour of the Lennard-Jones liquid.

% The model then has six fitting parameters: $a$, $b$, $c = \left[{\color{green}1-2\ln\Lambda}\right]$, $d$, $\gamma$ and {\color{green}$\phi=\left[\mu n_{\rm G} - \mu n_{\rm L} - f_{\rm G}\right]$}. This number can be reduced by noting that $a$ and $b$ are related by $a = b - B_2$, where $B_2$ is the second virial coefficient, which can be calculated for the truncated, shifted and smoothed Lennard-Jones potential on a sphere of a given radius and at a particular temperature.

\begin{figure}
\input{ld_7_z_sweep}
\caption{\label{fig:fitsGCZ} The free energy barrier to the nucleation of the liquid phase for Lennard-Jones particles on the surface of a sphere of radius $R=7$ in the grand canonical ensemble for an array of applied activities. The temperature for all simulations is $T = 0.40$. Solid lines correspond to simulation data and dashed lines to fits to the theoretical model described in Sec.~\ref{sub:nucGC}.}
\end{figure}
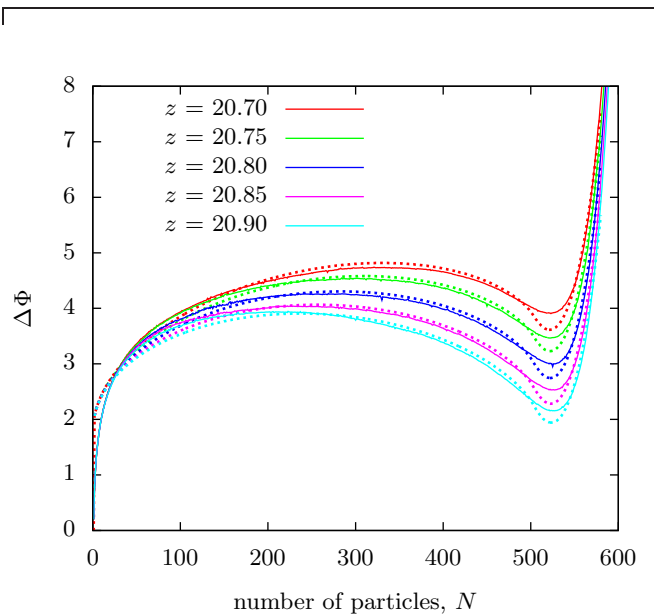

Fig.~\ref{fig:compGC} shows a fit of the model in Eq.~(\ref{equ:grandFull}) to the simulated nucleation curve at $z = 20.90$ and $T=0.40$. It can be seen that this model successfully captures the full free energy profile.  To demonstrate that we are not simply over-fitting the data, a series of simulations were performed with a range of applied activities, as shown in Fig.~\ref{fig:fitsGCZ}. We predict that the only fitting parameter that should be modified by changing the activity is $z_{\rm L}$ as this is related to the chemical potential of the out-of-equilibrium liquid droplet. Therefore, we can fit all the curves simultaneously, requiring that all fitting parameters, except $z_{\rm L}$, are the same for each curve. The values of the fitting parameters can be found in Table \ref{tab:gc}. It can be seen that even with these restrictions, the model fits all the data very well, suggesting that it does correctly describe all features of gas--liquid nucleation on a sphere in the grand canonical ensemble.  It is also worth noting that the fitted activity of the liquid domain $z_{\rm L}$ is close to that of the reservoir $z$. Furthermore, the liquid density $n_{\rm L}$ from the relation $A_{\rm L}(N)$ is constant at $0.87$ for all the curves in Fig.~\ref{fig:fitsGCZ}.

\begin{table}
\caption{\label{tab:gc} Fitting parameters for the grand canonical ensemble activity sweep shown in Fig.~\ref{fig:fitsGCZ}. The following fitting parameters were constrained to be identical across the whole set of simulations: $a=1.26$, $b=0.736$, $\gamma=0.054$, $d=1.78$.}
\begin{tabular}{c | c c c c c}
$\ln\left(z\right)$ & 3.033 & 3.035 & 3.037 & 3.040 & 3.042 \\
\hline
$\ln\left(z_{\rm L}\right)$ & 3.603 & 3.605 & 3.607 & 3.610 & 3.611\\
\end{tabular}
\end{table}

\subsection{\label{sub:GcCrit}Critical nuclei}

By comparing the geometric curvature of the perimeter of a spherical cap to that of a circle on a plane, G{\'o}mez {\it et al.} predicted that the radius of the critical nucleus depends on the radius of the spherical surface according to \cite{Gomez2015A}
\begin{equation}
     r_{\text{c}}(R) = R \arctan{\left(\frac{r_{\text{c}}^0}{R}\right)},\label{equ:gomez}
\end{equation}
where $r_{\text{c}}(R)$ is the (geodesic) radius of the critical nucleus on a sphere of radius $R$, given that the critical nucleus of the equivalent system on the plane would be $r_{\text{c}}^0$.
This mapping between the planar and spherical systems implicitly makes the assumption that the line tension and difference in free energy density between the phases are independent of the curvature of the surface.

\begin{figure}
\input{ld_r_sweep}
\caption{\label{fig:fitsGCR} Free energy profile measured for grand canonical systems on spheres of different radii, but with the chemical potential fixed so that it corresponds to an activity of $z_{\rm ref} = 20.90$ on a sphere of radius $R=7$. The temperature for all simulations is $T = 0.40$}. 
\end{figure}
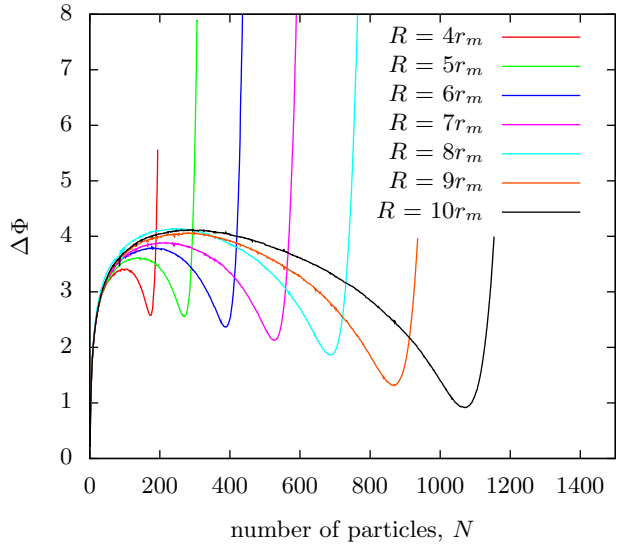

To test the prediction in Eq.~(\ref{equ:gomez}) over a range of conditions, we ran three sets of simulations. For each set, the temperature and chemical potential were kept constant while the radius of the confining sphere was varied.  For reference, we quote the activity $z_{\rm ref}$ corresponding to the chosen chemical potential at radius $R=7$, noting that Eq.~(\ref{zdef}) for the activity includes a factor of $A_0$.  The results for one such set of simulations are shown in Fig.~\ref{fig:fitsGCR}.
%\sout{The number of particles in the critical nucleus is obtained by fitting the simulation data with our model and locating the maximum. For these fits, all five fitting parameters were allowed to vary for each radius, since in principle they can vary with surface curvature.}

As discussed in the previous subsection, for each free energy profile, we typically observe a constant liquid density $n_{\rm L}$ until the liquid covers the sphere. We use this liquid density to convert the measured number of particles in the critical nucleus to the radius of the critical nucleus, $r_{\text{c}}$, assuming that the critical nucleus is a spherical cap.
We estimate the uncertainty in $r_{\rm c}$ by repeating the simulation of a representative free energy curve five times, then calculating the uncertainty in the barrier height. By comparing this to the form of the nucleation curve at the top of the barrier, we estimate the error in the number of particles in a critical nucleus to be approximately $\pm6$ and from this calculate the corresponding error in the radius.
 Each set of simulations can then be  fitted to Eq.~(\ref{equ:gomez}) to obtain $r_{\text{c}}^0$, the corresponding critical nucleus radius on the flat surface.  We note that, in several cases, the predicted value of $r_{\text{c}}^0$ would be impractically large to measure in a simulation.
% We are unable to directly measure $r_{\text{c}}^0$ because it is usually so large that it would be impossible to simulate a system large enough to contain the critical nucleus in a reasonable time. 

\begin{figure}
\input{crit_radius}
\caption{\label{fig:gcCrit} The size of the critical nucleus varies with the radius of the confining sphere. These data are plotted for three sets of grand canonical simulations at different temperatures and activities.}
\end{figure}
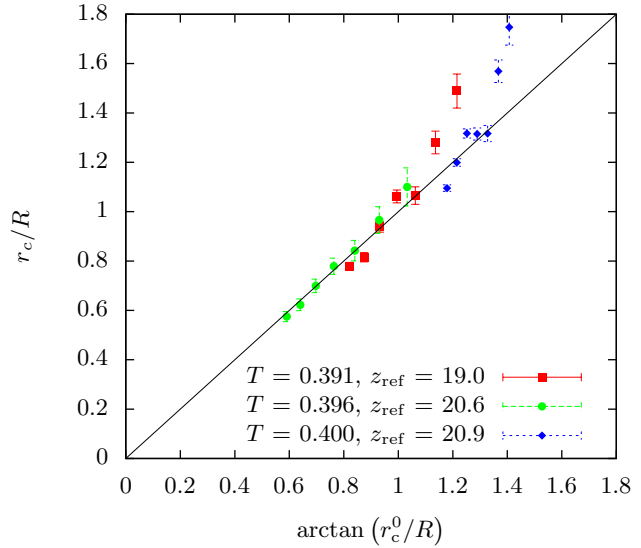

\begin{table}
\caption{Line tensions extracted from the free energy profiles using Eq.~(\ref{equ:grandFull}) for the points plotted in Fig.~\ref{fig:gcCrit}. The top row corresponds to the red squares in Fig.~\ref{fig:gcCrit}. The second row corresponds to green circles and the third to blue diamonds.\label{tab:gamma}}
\begin{tabular}{c | c c c c c c c}
\hline
$(z_{\rm ref},T)$ & $R = 4$ & $R = 5$ & $R = 6$ & $R = 7$ & $R = 8$ & $R = 9$ & $R =10$ \\
\hline
(19.0,0.391) & 0.096 & 0.086 & 0.079 & 0.077 & 0.073 & 0.071 & 0.077 \\
(20.6,0.396) & 0.088 & 0.079 & 0.075 & 0.077 & 0.076 & 0.075 & 0.072 \\
(20.9,0.400) & 0.068 & 0.058 & 0.054 & 0.049 & 0.049 & 0.044 & 0.041 \\
\end{tabular}
\end{table}

For one activity--temperature pair $(z_{\rm ref},T)$, the simulation results can be fitted well to Eq.~(\ref{equ:gomez}), and they fall close to a single line as shown in Fig.~\ref{fig:gcCrit} (green circles). However, for the remaining two $(z_{\rm ref},T)$ pairs, the fit breaks down for the points which correspond to two smallest radii: $R = 4$ and $R = 5$. To understand why this is the case, we fit Eq.~(\ref{equ:grandFull}) to the free energy profiles to extract the line tension in each case. For one of the sets (green circles), the line tension is almost independent of sphere radius, as shown in Table \ref{tab:gamma}. In this set of data, the geometrical correction embodied in Eq.~(\ref{equ:gomez}) is sufficient to relate the spherical and planar cases. In contrast, for the remaining sets (red squares and blue diamonds), which correspond to lower supersaturations, the fitted line tension decreases with increasing radius of curvature (Table \ref{tab:gamma}), and this effect is particularly pronounced at low radii. This trend could reflect either increased sensitivity of the surface tension to curvature as the phase boundary is approached, or greater deviation from the idealised spherical cap shape that the nucleus is assumed to adopt.  Eq.~(\ref{equ:gomez}) neglects any curvature dependence of the line tension or nucleus shape \cite{Gomez2015A}, and Fig.~\ref{fig:gcCrit} demonstrates that such dependence can have a significant influence.

%For two activity--temperature pairs ($z_{\rm ref}$, $T$), the simulation results can be fitted well to Eq.~(\ref{equ:gomez}), and they fall close to a single line as shown in Fig.~\ref{fig:gcCrit} (blue circles and green diamonds). However, for $T = 0.481$ and $z_{\rm ref} = 20.9$ (red squares), Eq.~(\ref{equ:gomez}) is a poor fit. To understand why this is the case, we fit Eq.~(\ref{equ:grandFull}) to the free energy profiles to extract the line tension in each case. For two of the sets (blue circles and green diamonds), the line tension is almost independent of sphere radius, as shown in Table \ref{tab:gamma}.  In these sets of data, the geometrical correction embodied in Eq.~(\ref{equ:gomez}) is sufficient to relate the spherical and planar cases.  In contrast, for the remaining set (red squares), which corresponds to the highest temperature studied, the fitted line tension decreases significantly with increasing radius of curvature (Table \ref{tab:gamma}).  This trend could reflect either increased sensitivity of the surface tension to curvature as the critical point is approached, or greater deviation from the idealised spherical cap shape that the nucleus is assumed to adopt.  Eq.~(\ref{equ:gomez}) neglects any curvature dependence of the line tension or nucleus shape \cite{Gomez2015A}, and Fig.~\ref{fig:gcCrit} demonstrates that such dependence can have a significant influence.

\section{\label{sec:canonical} The Canonical Ensemble}

\begin{figure}
\includegraphics[width=0.48\textwidth]{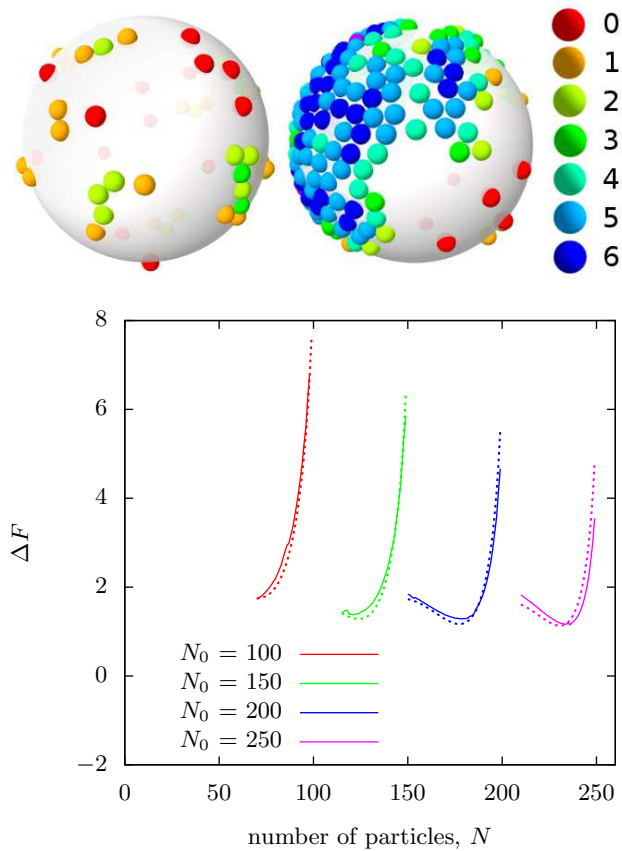}
\input{ld_canon}
\caption{\label{fig:canonical} Top: Snapshots of equilibrium configurations taken from simulations on a sphere of radius $R = 7$ at $T = 0.40$ in the canonical ensemble. At $N_0 = 50$ (left) the surface is covered by a homogeneous gas of isolated particles and small clusters. At $N_0 = 250$ (right) there is a single cap-shaped liquid drop in equilibrium with a gas.  Particles are coloured according to their coordination number as indicated in the legend.  Bottom: The free energy variation of a liquid cluster around its equilibrium size on a sphere of radius $R=7$ for four quantities of particles in the canonical ensemble. The temperature for all simulations is $T = 0.40$. Solid lines are taken from umbrella sampling simulations, while the dashed lines are fits to the MRB model. The values of the parameters of the model can be found in Table \ref{tab:c}.}
\end{figure}

In the canonical ensemble, the total number of particles, $N_0$, is fixed in a given system.
As the spherical surface we are considering is intrinsically finite in extent, the canonical and grand canonical ensembles are not equivalent. In particular, the density of the gas phase decreases when a liquid droplet forms in the canonical ensemble because the gas is not replenished by a particle bath as in the grand canonical ensemble.

Similar considerations apply to three-dimensional \cite{Schrader09a} and planar two-dimensional \cite{Rovere1990} systems with periodic boundary conditions in the canonical ensemble.  In these cases, an evaporation--condensation transition is observed as a function of increasing density.  First, a droplet condenses from the supersaturated vapour.  At higher densities the condensed phase grows and a percolating slab geometry becomes preferable; this geometry reduces the surface area or perimeter of the condensed phase and is effectively stabilised by the periodic boundary conditions.  Finally, for even denser cases, the vapour phase is reduced to a bubble within a continuous condensed phase.  In the spherically confined case, we observe states analogous to the initial gas state and to the droplet--gas equilibrium, as illustrated by the snapshots in Fig.~\ref{fig:canonical}.  The bubble state is continuously reached from the droplet by increasing the density, as the droplet simply grows to cover more than half of the sphere's surface.  The equivalent of the slab phase would be a condensed band, or spherical cap with a hole.  This intermediate state does not arise on the sphere since, unlike the periodic planar case, the creation of the second interface always increases the total perimeter of the droplet and is therefore thermodynamically unfavourable.

We have used umbrella sampling simulations to follow the free energy of a liquid droplet in the canonical ensemble for a selection of values of $N_0$.
Fig.~\ref{fig:canonical} shows the free energy of a system containing a single cluster embedded in the gas phase, around the equilibrium cluster size. In this case, the sharp increase in free energy with the size of the liquid cluster is not caused by high density across the surface. Instead, at large cluster sizes, the gas phase is depleted below its equilibrium density. As more gas particles are absorbed into the liquid cluster, the loss of entropy outweighs the energy gain of joining the liquid cluster. In contrast to the grand canonical ensemble, the position of the minimum is heavily dependent on the initial density of the parent gas phase in the canonical ensemble.

For sufficiently small $N_0$ (in the vicinity of 100 in Fig.~\ref{fig:canonical}), the sphere is uniformly covered in a gas of isolated particles and small clusters, and the local minimum in the free energy corresponding to the droplet vanishes.  Although it is hard to precisely locate the value of $N_0$ at which the minimum disappears, this point, in principle, defines the density of the finite-size equivalent of the first-order phase transition.  This threshold is an intrinsic property of a given sphere size and depends on its radius, but should approach the thermodynamic limit of the planar case as the radius of the sphere becomes very large.

In order to understand the free energy curves, we have developed a model for the free energy of a single liquid cluster surrounded by gas and confined to a sphere in the canonical ensemble. This model is based on one developed by Rao and Berne for finite, Euclidean, three-dimensional systems in the canonical ensemble~\cite{Rao1979A}. A related approach was also adopted by Reguera {\it et al.}~\cite{Reguera2003}. The model is based on the assumption that a single liquid cluster grows in the finite bath of the gas phase. Our version of the model accounts for the two-dimensional spherical geometry by introducing the line tension term described in Eq.~(\ref{linetension}).
As in the original model by Rao and Berne~\cite{Rao1979A}, we choose to truncate the virial expansion at the second term. The model is derived in detail in Appendix \ref{apx:roa} and its complete form is
\begin{widetext}
\begin{gather}
	\Delta F = N_{\text{0}}k_{\rm B}T \left[\ln{\left(\frac{n_{\text{G}}}{n_{\text{0}}}\right)} + 2 B_2 \left(n_{\text{G}} - n_{\text{0}}\right)\right] + Nk_{\rm B}T \left[\ln{\left(\frac{n_{\text{G$\infty$}}}{n_{\text{G}}}\right)} + 2 B_2 \left(n_{\text{G$\infty$}} - n_{\text{G}}\right) + \frac{n_{\text{G}} - n_{\text{G$\infty$}} + B_2 \left( n_{\text{G}}^2 - n_{\text{G$\infty$}}^2 \right)}{{ n_{\text{L$\infty$}}}}\right] \nonumber \\
- A_{\text{0}} k_{\rm B}T \left[n_{\text{G}} + B_2 n_{\text{G}}^2 - n_{\text{0}} - B_2 n_{\text{0}}^2\right] + \gamma \sqrt{4\pi\left(\frac{N}{n_{\text{L$\infty$}}} - \frac{N^2}{n_{\text{L$\infty$}}A_{\text{0}}}\right)} + d, 
 \label{equ:berne}
\end{gather}
\end{widetext}
where
\begin{equation}
	n_{\text{G}} = \frac{N_{\text{0}} - N}{A_{\text{0}} - N/n_{\text{L$\infty$}}}, \label{equ:ng}
\end{equation}
is the density of the gas in the space not occupied by the liquid cluster.
$N_{\text{0}}$ is the total number of particles on the sphere of area $A_{0}$, giving an overall density $n_0=N_0/A_0$. $B_2$ is the second virial coefficient, which depends on the curvature of the surface and is evaluated by numerical integration. The parameters in the model are $n_{\text{L$\infty$}}$, the equilibrium liquid density; $n_{\text{G$\infty$}}$, the equilibrium gas density; $\gamma$, the line tension; and $d$, a constant which contains the translational entropy of the cluster. All other symbols have their previous meanings. We refer to Eq.~(\ref{equ:berne}) as the modified Rao--Berne (MRB) model.

When applying the MRB model, we take the line tension $\gamma$ obtained in the grand canonical simulations in Sec.~\ref{sec:GrandCanonical} with the same sphere radius and temperature. The equilibrium liquid density, $n_{\text{L$\infty$}}$, is estimated as follows. A simulation is run without umbrella sampling. During the simulation, the total number of particles in a liquid-like environment (as defined in Sec.~\ref{sub:reactCoord}) is averaged. The Voronoi decomposition on the sphere is also taken of the particle locations~\cite{Aurenhammer2013A}. A histogram of the area of each Voronoi cell is constructed. The location of the peak of this histogram is an estimate for the inverse density of the liquid phase, as most particles are found in the bulk of the liquid cluster. 
%The density of the gas phase can then be estimated using
%
%\begin{displaymath}
 %    n_{\text{G} \infty} \approx \frac{N_{\text{0}} - {\bar N}_{\text{L}}}{A_{\text{0}} - {\bar N}_{\text{L}} n_{\text{L} \infty}},
%\end{displaymath} 
%
% where ${\bar N}_{\text{L}}$ is the average number of particles in a liquid-like environment.

This leaves only the equilibrium reference gas density $n_{\text{G$\infty$}}$ and the offset $d$ to be fitted. We constrain $n_{\text{G$\infty$}}$ to be the same for all runs, since it should be the same for a given temperature and sphere radius, and we allow $d$ to be different for each curve. The results of these fits to the simulation data are shown as dashed lines in Fig.~\ref{fig:canonical} and the corresponding values of the parameters are given in Table \ref{tab:c}. It can be seen that the theory fits the simulation closely, for systems with a single large cluster.

\begin{table}
\caption{\label{tab:c}The calculated and fitted parameters for liquid cluster formation on a sphere of radius $R=7$ at temperature $T=0.40$ in the canonical ensemble. $n_{\rm L\infty}$, $B_2$ and $\gamma$ are calculated and have the values of $n_{\rm L \infty} = 0.951$, $B_2 = -6.247$, $\gamma=0.054$. $n_{\rm G \infty}$ and $d$ are fitted. $n_{\rm G \infty}$ is constrained to have the same value for all runs, and the best fit value is 0.120. $d$ is fitted separately for each run.}
\begin{tabular}{c | c}
$N_0$  & $d$\\
\hline
$100$ & $-0.031$\\
$150$ & $-11.40$\\
$200$ & $-35.79$\\
$250$ & $-75.23$\\    
\end{tabular}
\end{table}

\section{\label{sec:conclusions}Conclusions}

We have shown that the thermodynamics of a two-dimensional soft system in disordered phases confined to the surface of a sphere is dominated by the effects of the surface geometry on the cluster boundary and on the finite extent of the surface. Unlike the planar two-dimensional case, the finite size of these curved systems is a natural consequence of being embedded on a sphere, providing a physical realisation of an intrinsically finite system with neither hard boundaries nor artificial periodic boundary conditions.

In the grand canonical ensemble, the phase diagram of a system of Lennard-Jones particles on a sphere is indistinguishable from that of a periodic plane of the same area. This is because both phases are fluids and neither of the two (meta-)stable states of the system contains a phase boundary; the sphere is fully covered in either liquid or gas. However, the nucleation process is influenced by curvature. The free energy of the growing nucleus is initially modified by the geometry of the surface, and growth is later arrested when the surface is fully covered. A model, based on classical nucleation theory and modified to handle these additional considerations, is able to rationalise the simulation results. By investigating spheres of different radii over a range of conditions, we were also able to confirm the prediction for the curvature dependence of the critical nucleus size made by G{\'o}mez {\it et al.}~\cite{Gomez2015A} provided the sphere radius and supersaturation is not too low.  At low supersaturation, we find that the line tension decreases monotonically with increasing radius of curvature, which conflicts with the assumption of constant line tension used in Ref.~\cite{Gomez2015A}.

Due to the finite size of the system, ensemble equivalence does not hold for the system we are considering. As discussed in Sec.~\ref{sub:umbrella} nucleation is not observed in the canonical ensemble of the Lennard-Jones particles. Instead, at sufficiently high densities a stable liquid cluster is established in equilibrium with a depleted gas phase, and the free energy for formation of the droplet at around its equilibrium size could be measured. In contrast to the analogous evaporation--condensation transition in a planar two-dimensional system \cite{Rovere1990}, there is no intermediate slab-like phase because creation of a second gas--liquid interface can never reduce the total perimeter of a droplet on a sphere. A modified Rao--Berne model (with just two fitting parameters for a whole family of free energy profiles) was able to capture both the equilibrium liquid cluster size and the free energy of clusters close to the equilibrium size.

There are a number of possible extensions to the work presented here. Firstly, we have concentrated on the disordered phases. An interesting future direction is to consider the nucleation and/or growth of the crystalline phases. Secondly, it is now possible to experimentally realise surfaces with non-constant Gaussian curvature, such as tori and unduloids~\cite{Kusumaatmaja2013A,Irvine2010A,Pairam2014A}. In contrast to spherical surfaces, curvature inhomogeneities on these surfaces break the translational invariance of the nucleation process. Thirdly, many real two dimensional systems are confined to flexible, rather than fixed, curved surfaces \cite{Kabaso2011A,Dias200A,Abel2018A,Lipowsky1997A}. Examining the interdependence of the confined colloidal phase and the shape of the underlying confining surface is a challenging and important open problem.

\section{Acknowledgements}

The authors thank Colin Bain, Mike Evans, Johan Mattsson, Stefan Paquay and Patrick Warren for valuable comments and discussions.
J.O.L.~is grateful to the Engineering and Physical Sciences Research Council (EPSRC) Centre for Doctoral Training in Soft Matter and Functional Interfaces (grant EP/L015536/1) for financial support.

\section{Open data}

The simulation code, sample scripts and all data plotted in this work are available at: https://doi.org/10.15128/r1jh343s29p.

\appendix
\section{\label{apx:roa}The modified Rao--Berne model}

Rao and Berne formulated a model of nucleation in finite three-dimensional systems at constant $NVT$~\cite{Rao1979A}. The derivation below closely follows theirs, but we consider the two dimensional case. Up to the second virial coefficient, the vapour pressure $p$ as a function of the density $n_{\rm G}$ is
\begin{equation} 
	p(n_{\rm G})= n_{\text{G}}k_{\text{B}}T \left( 1 + B_{2} n_{\text{G}} \right),
\label{equ:p}
\end{equation}  
where  $T$ is the temperature and $n_{\text{G}}$ is given by Eq.~(\ref{equ:ng}). The chemical potential per molecule ${\mu}_{\text{G}}$ is 
\begin{equation}
	{\mu}_{\text{G}}(n_{\rm G}) = {\mu}_{\text{G}}^{0} + k_{\text{B}}T \left( \ln n_{\text{G}} + 2 B_2 n_{\text{G}} \right), \label{equ:cp}
\end{equation}
where ${\mu}_{\text{G}}^{0}$ is a reference chemical potential. The total Gibbs free energy $G$ of a system containing a droplet surrounded by gas is
\begin{displaymath}
	G=G_{\text{G}} + G_{\text{L}} + G_{\text{int}},
\end{displaymath}
where
\begin{equation}
	G_{\text{G}} = (N_{\text{0}} - N) {\mu}_{\text{G}},
\label{equ:Ggas}
\end{equation}
\begin{equation}
	G_{\text{L}} = N {\mu}_{\text{L}},
\label{equ:Gliquid}
\end{equation}
and
\begin{equation}
	G_{\text{int}} = \gamma P.
\label{equ:Gperim}
\end{equation}
$N_{\text{0}}$ is the total number of particles in the system, $N$ is the number of particles in the liquid cluster, ${\mu}_{\text{L}}$ is the chemical potential per liquid particle and $P$ is the perimeter of the cluster. ${\mu}_{\text{L}}$ can be calculated using the Gibbs-Duhem relation
\begin{equation}
	{\mu}_{\text{L}} = {\mu}_{\text{L}}\left( n_{\text{L$\infty$}} \right) + \frac{1}{n_{\text{L$\infty$}}}\left( p - p_{\text{$\infty$}} \right),
\label{equ:muL}
\end{equation}
where $n_{\text{L$\infty$}}$ is the equilibrium liquid density and $p_{\text{$\infty$}}$ is the corresponding reference pressure. Noting that at equilibrium
\begin{equation}
	{\mu}_{\text{L}}\left( n_{\text{L$\infty$}} \right) = {\mu}_{\text{G}}\left( n_{\text{G$\infty$}} \right),
\label{equ:muLmuG}
\end{equation}
${\mu}_{\text{G}}\left( n_{\text{G$\infty$}} \right)$ and hence ${\mu}_{\text{L}}\left( n_{\text{L$\infty$}} \right)$ can be calculated from Eq.~(\ref{equ:cp}). Using Eq. \eqref{equ:p}, we can also write the equilibrium pressure as 
\begin{equation} 
	p_{\text{$\infty$}}= n_{\text{G$\infty$}}k_{\text{B}}T \left( 1 + B_{2} n_{\text{G$\infty$}} \right).
\label{equ:peq}
\end{equation}  

The Helmholtz free energy $F$ in two dimensions is 
\begin{displaymath}
	F = G - pA.
\end{displaymath}
Starting from a gas of initial density $n_0=N_0/A_0$, the free energy change associated with the formation of a liquid cluster containing $N$ particles is 
\begin{equation}
	\Delta F = G - N_0 {\mu}_{\text{G}}(n_0) - \left[p(n_{\rm G}) - p(n_0) \right] A_0.
\label{equ:DeltaF}
\end{equation}
\par
We now have all the components required to assemble Eq.~(\ref{equ:berne}).  Inserting Eqs.~(\ref{equ:Ggas}) to (\ref{equ:Gperim}) into Eq.~(\ref{equ:DeltaF}) and collecting terms gives
\begin{displaymath}
\begin{split}
\Delta F = N_0\left[\mu_{\rm G}(n_{\rm G}) - \mu_{\rm G}(n_0)\right] + N\left[\mu_{\rm L}-\mu_{\rm G}(n_{\rm G})\right] - \\
\left[p(n_{\rm G})-p(n_0)\right]A_0 + \gamma P.
\end{split}
\end{displaymath}
Substitution of Eqs.~(\ref{linetension}), (\ref{equ:p}), (\ref{equ:cp}), (\ref{equ:muL}) and (\ref{equ:peq}) for $P$, $p$, $\mu_{\rm G}$, $\mu_{\rm L}$ and $p_{\text{$\infty$}}$ then gives the full expression in Eq.~(\ref{equ:berne}).
\par
In this approach to the canonical ensemble, the chemical potentials of the reference liquid and gas phases have been directly equated in Eq.~(\ref{equ:muLmuG}) and it is necessary to account for the non-ideality of the gas (via the second virial coefficient) to obtain acceptable results.  In contrast, for the treatment of the grand canonical ensemble in Sec.~\ref{sec:GrandCanonical}, we approach the liquid and gas contributions explicitly, using van der Waals-like parameters to describe the liquid. There, it is sufficient to treat the gas as ideal, and including the non-ideality of the gas results in negligible changes to the fitting parameters.

\bibliographystyle{apsrev4-1}
\bibliography{rsc}

\clearpage
\begin{figure}
\includegraphics[width=1.0\textwidth]{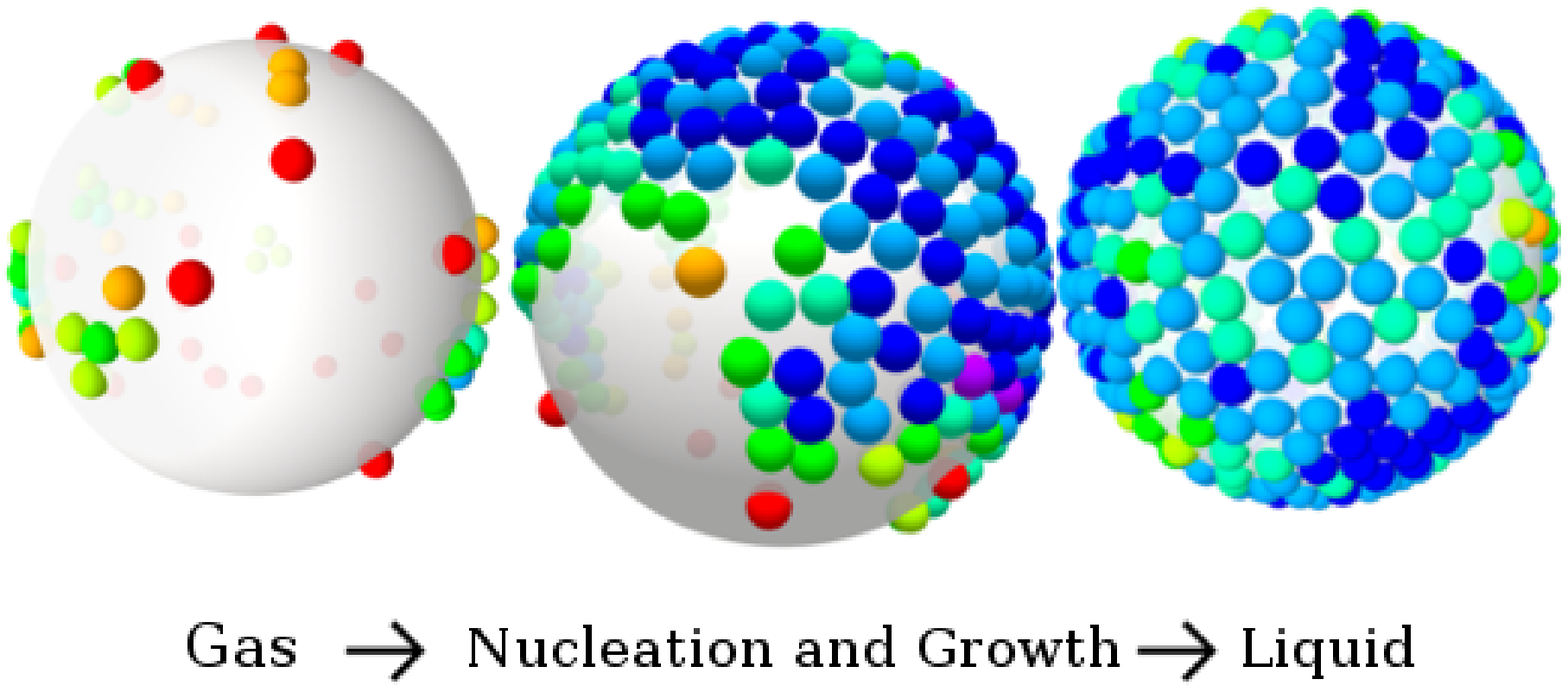}
\end{figure}
\end{document}

%% file: phase_diag.tex
% GNUPLOT: LaTeX picture with Postscript
\begingroup
  \makeatletter
  \providecommand\color[2][]{%
    \GenericError{(gnuplot) \space\space\space\@spaces}{%
      Package color not loaded in conjunction with
      terminal option `colourtext'%
    }{See the gnuplot documentation for explanation.%
    }{Either use 'blacktext' in gnuplot or load the package
      color.sty in LaTeX.}%
    \renewcommand\color[2][]{}%
  }%
  \providecommand\includegraphics[2][]{%
    \GenericError{(gnuplot) \space\space\space\@spaces}{%
      Package graphicx or graphics not loaded%
    }{See the gnuplot documentation for explanation.%
    }{The gnuplot epslatex terminal needs graphicx.sty or graphics.sty.}%
    \renewcommand\includegraphics[2][]{}%
  }%
  \providecommand\rotatebox[2]{#2}%
  \@ifundefined{ifGPcolor}{%
    \newif\ifGPcolor
    \GPcolortrue
  }{}%
  \@ifundefined{ifGPblacktext}{%
    \newif\ifGPblacktext
    \GPblacktextfalse
  }{}%
  % define a \g@addto@macro without @ in the name:
  \let\gplgaddtomacro\g@addto@macro
  % define empty templates for all commands taking text:
  \gdef\gplbacktext{}%
  \gdef\gplfronttext{}%
  \makeatother
  \ifGPblacktext
    % no textcolor at all
    \def\colorrgb#1{}%
    \def\colorgray#1{}%
  \else
    % gray or color?
    \ifGPcolor
      \def\colorrgb#1{\color[rgb]{#1}}%
      \def\colorgray#1{\color[gray]{#1}}%
      \expandafter\def\csname LTw\endcsname{\color{white}}%
      \expandafter\def\csname LTb\endcsname{\color{black}}%
      \expandafter\def\csname LTa\endcsname{\color{black}}%
      \expandafter\def\csname LT0\endcsname{\color[rgb]{1,0,0}}%
      \expandafter\def\csname LT1\endcsname{\color[rgb]{0,1,0}}%
      \expandafter\def\csname LT2\endcsname{\color[rgb]{0,0,1}}%
      \expandafter\def\csname LT3\endcsname{\color[rgb]{1,0,1}}%
      \expandafter\def\csname LT4\endcsname{\color[rgb]{0,1,1}}%
      \expandafter\def\csname LT5\endcsname{\color[rgb]{1,1,0}}%
      \expandafter\def\csname LT6\endcsname{\color[rgb]{0,0,0}}%
      \expandafter\def\csname LT7\endcsname{\color[rgb]{1,0.3,0}}%
      \expandafter\def\csname LT8\endcsname{\color[rgb]{0.5,0.5,0.5}}%
    \else
      % gray
      \def\colorrgb#1{\color{black}}%
      \def\colorgray#1{\color[gray]{#1}}%
      \expandafter\def\csname LTw\endcsname{\color{white}}%
      \expandafter\def\csname LTb\endcsname{\color{black}}%
      \expandafter\def\csname LTa\endcsname{\color{black}}%
      \expandafter\def\csname LT0\endcsname{\color{black}}%
      \expandafter\def\csname LT1\endcsname{\color{black}}%
      \expandafter\def\csname LT2\endcsname{\color{black}}%
      \expandafter\def\csname LT3\endcsname{\color{black}}%
      \expandafter\def\csname LT4\endcsname{\color{black}}%
      \expandafter\def\csname LT5\endcsname{\color{black}}%
      \expandafter\def\csname LT6\endcsname{\color{black}}%
      \expandafter\def\csname LT7\endcsname{\color{black}}%
      \expandafter\def\csname LT8\endcsname{\color{black}}%
    \fi
  \fi
  \setlength{\unitlength}{0.0500bp}%
  \begin{picture}(5040.00,4320.00)%
    \gplgaddtomacro\gplbacktext{%
      \csname LTb\endcsname%
      \put(726,704){\makebox(0,0)[r]{\strut{}$0.34$}}%
      \put(726,1542){\makebox(0,0)[r]{\strut{}$0.36$}}%
      \put(726,2380){\makebox(0,0)[r]{\strut{}$0.38$}}%
      \put(726,3217){\makebox(0,0)[r]{\strut{}$0.4$}}%
      \put(726,4055){\makebox(0,0)[r]{\strut{}$0.42$}}%
      \put(858,484){\makebox(0,0){\strut{}$0$}}%
      \put(1331,484){\makebox(0,0){\strut{}$0.1$}}%
      \put(1804,484){\makebox(0,0){\strut{}$0.2$}}%
      \put(2277,484){\makebox(0,0){\strut{}$0.3$}}%
      \put(2751,484){\makebox(0,0){\strut{}$0.4$}}%
      \put(3224,484){\makebox(0,0){\strut{}$0.5$}}%
      \put(3697,484){\makebox(0,0){\strut{}$0.6$}}%
      \put(4170,484){\makebox(0,0){\strut{}$0.7$}}%
      \put(4643,484){\makebox(0,0){\strut{}$0.8$}}%
      \put(220,2379){\rotatebox{-270}{\makebox(0,0){\strut{}$T$}}}%
      \put(2750,154){\makebox(0,0){\strut{}number density, $n$}}%
    }%
    \gplgaddtomacro\gplfronttext{%
      \csname LTb\endcsname%
      \put(2719,2489){\makebox(0,0)[r]{\strut{}sphere}}%
      \csname LTb\endcsname%
      \put(2719,2269){\makebox(0,0)[r]{\strut{}plane}}%
    }%
    \gplbacktext
    \put(0,0){\includegraphics{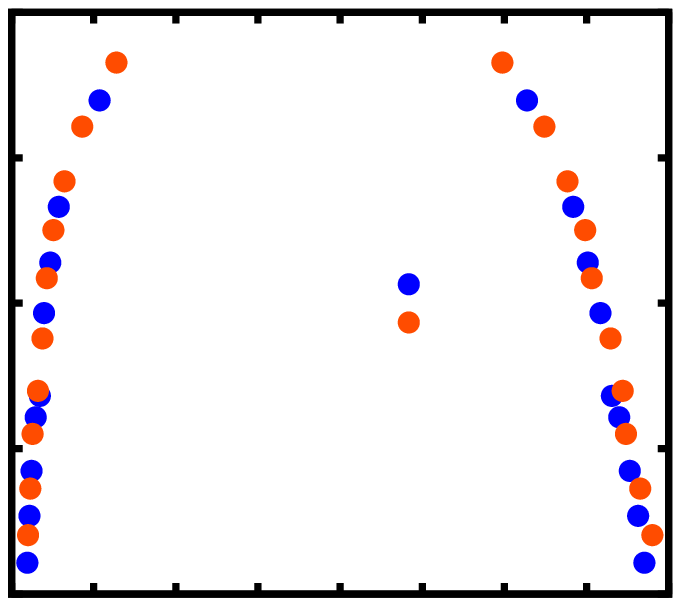}}%
    \gplfronttext
  \end{picture}%
\endgroup

%% file: ld_compare.tex
% GNUPLOT: LaTeX picture with Postscript
\begingroup
  \makeatletter
  \providecommand\color[2][]{%
    \GenericError{(gnuplot) \space\space\space\@spaces}{%
      Package color not loaded in conjunction with
      terminal option `colourtext'%
    }{See the gnuplot documentation for explanation.%
    }{Either use 'blacktext' in gnuplot or load the package
      color.sty in LaTeX.}%
    \renewcommand\color[2][]{}%
  }%
  \providecommand\includegraphics[2][]{%
    \GenericError{(gnuplot) \space\space\space\@spaces}{%
      Package graphicx or graphics not loaded%
    }{See the gnuplot documentation for explanation.%
    }{The gnuplot epslatex terminal needs graphicx.sty or graphics.sty.}%
    \renewcommand\includegraphics[2][]{}%
  }%
  \providecommand\rotatebox[2]{#2}%
  \@ifundefined{ifGPcolor}{%
    \newif\ifGPcolor
    \GPcolortrue
  }{}%
  \@ifundefined{ifGPblacktext}{%
    \newif\ifGPblacktext
    \GPblacktextfalse
  }{}%
  % define a \g@addto@macro without @ in the name:
  \let\gplgaddtomacro\g@addto@macro
  % define empty templates for all commands taking text:
  \gdef\gplbacktext{}%
  \gdef\gplfronttext{}%
  \makeatother
  \ifGPblacktext
    % no textcolor at all
    \def\colorrgb#1{}%
    \def\colorgray#1{}%
  \else
    % gray or color?
    \ifGPcolor
      \def\colorrgb#1{\color[rgb]{#1}}%
      \def\colorgray#1{\color[gray]{#1}}%
      \expandafter\def\csname LTw\endcsname{\color{white}}%
      \expandafter\def\csname LTb\endcsname{\color{black}}%
      \expandafter\def\csname LTa\endcsname{\color{black}}%
      \expandafter\def\csname LT0\endcsname{\color[rgb]{1,0,0}}%
      \expandafter\def\csname LT1\endcsname{\color[rgb]{0,1,0}}%
      \expandafter\def\csname LT2\endcsname{\color[rgb]{0,0,1}}%
      \expandafter\def\csname LT3\endcsname{\color[rgb]{1,0,1}}%
      \expandafter\def\csname LT4\endcsname{\color[rgb]{0,1,1}}%
      \expandafter\def\csname LT5\endcsname{\color[rgb]{1,1,0}}%
      \expandafter\def\csname LT6\endcsname{\color[rgb]{0,0,0}}%
      \expandafter\def\csname LT7\endcsname{\color[rgb]{1,0.3,0}}%
      \expandafter\def\csname LT8\endcsname{\color[rgb]{0.5,0.5,0.5}}%
    \else
      % gray
      \def\colorrgb#1{\color{black}}%
      \def\colorgray#1{\color[gray]{#1}}%
      \expandafter\def\csname LTw\endcsname{\color{white}}%
      \expandafter\def\csname LTb\endcsname{\color{black}}%
      \expandafter\def\csname LTa\endcsname{\color{black}}%
      \expandafter\def\csname LT0\endcsname{\color{black}}%
      \expandafter\def\csname LT1\endcsname{\color{black}}%
      \expandafter\def\csname LT2\endcsname{\color{black}}%
      \expandafter\def\csname LT3\endcsname{\color{black}}%
      \expandafter\def\csname LT4\endcsname{\color{black}}%
      \expandafter\def\csname LT5\endcsname{\color{black}}%
      \expandafter\def\csname LT6\endcsname{\color{black}}%
      \expandafter\def\csname LT7\endcsname{\color{black}}%
      \expandafter\def\csname LT8\endcsname{\color{black}}%
    \fi
  \fi
  \setlength{\unitlength}{0.0500bp}%
  \begin{picture}(5040.00,4320.00)%
    \gplgaddtomacro\gplbacktext{%
      \csname LTb\endcsname%
      \put(550,704){\makebox(0,0)[r]{\strut{}$0$}}%
      \put(550,1263){\makebox(0,0)[r]{\strut{}$1$}}%
      \put(550,1821){\makebox(0,0)[r]{\strut{}$2$}}%
      \put(550,2380){\makebox(0,0)[r]{\strut{}$3$}}%
      \put(550,2938){\makebox(0,0)[r]{\strut{}$4$}}%
      \put(550,3497){\makebox(0,0)[r]{\strut{}$5$}}%
      \put(550,4055){\makebox(0,0)[r]{\strut{}$6$}}%
      \put(682,484){\makebox(0,0){\strut{}$0$}}%
      \put(1342,484){\makebox(0,0){\strut{}$100$}}%
      \put(2002,484){\makebox(0,0){\strut{}$200$}}%
      \put(2663,484){\makebox(0,0){\strut{}$300$}}%
      \put(3323,484){\makebox(0,0){\strut{}$400$}}%
      \put(3983,484){\makebox(0,0){\strut{}$500$}}%
      \put(4643,484){\makebox(0,0){\strut{}$600$}}%
      \put(176,2379){\rotatebox{-270}{\makebox(0,0){\strut{}$\Delta \Phi$}}}%
      \put(2662,154){\makebox(0,0){\strut{}number of particles, $N$}}%
    }%
    \gplgaddtomacro\gplfronttext{%
      \csname LTb\endcsname%
      \put(3656,3882){\makebox(0,0)[r]{\strut{}simulation}}%
      \csname LTb\endcsname%
      \put(3656,3662){\makebox(0,0)[r]{\strut{}CNT planar model}}%
      \csname LTb\endcsname%
      \put(3656,3442){\makebox(0,0)[r]{\strut{}G{\'o}mez model {\it et al.} \cite{Gomez2015A} }}%
      \csname LTb\endcsname%
      \put(3656,3222){\makebox(0,0)[r]{\strut{}spherical model}}%
    }%
    \gplbacktext
    \put(0,0){\includegraphics{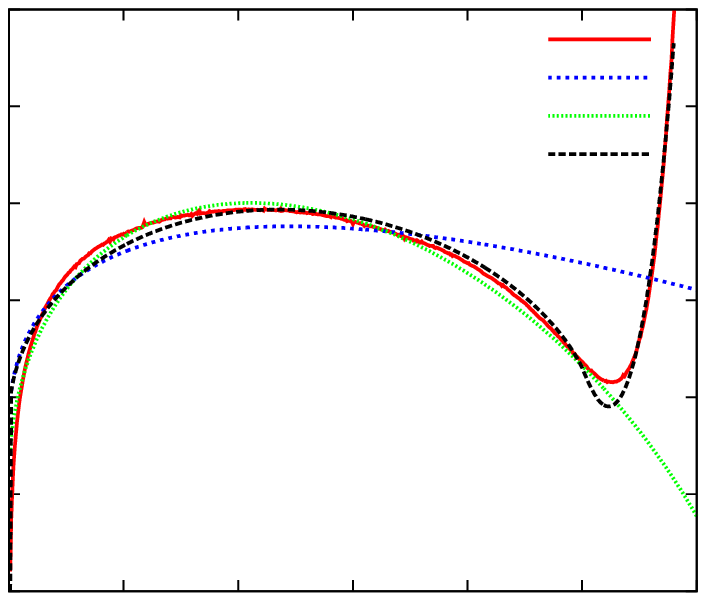}}%
    \gplfronttext
  \end{picture}%
\endgroup

%% file: ld_7_z_sweep.tex
% GNUPLOT: LaTeX picture with Postscript
\begingroup
  \makeatletter
  \providecommand\color[2][]{%
    \GenericError{(gnuplot) \space\space\space\@spaces}{%
      Package color not loaded in conjunction with
      terminal option `colourtext'%
    }{See the gnuplot documentation for explanation.%
    }{Either use 'blacktext' in gnuplot or load the package
      color.sty in LaTeX.}%
    \renewcommand\color[2][]{}%
  }%
  \providecommand\includegraphics[2][]{%
    \GenericError{(gnuplot) \space\space\space\@spaces}{%
      Package graphicx or graphics not loaded%
    }{See the gnuplot documentation for explanation.%
    }{The gnuplot epslatex terminal needs graphicx.sty or graphics.sty.}%
    \renewcommand\includegraphics[2][]{}%
  }%
  \providecommand\rotatebox[2]{#2}%
  \@ifundefined{ifGPcolor}{%
    \newif\ifGPcolor
    \GPcolortrue
  }{}%
  \@ifundefined{ifGPblacktext}{%
    \newif\ifGPblacktext
    \GPblacktextfalse
  }{}%
  % define a \g@addto@macro without @ in the name:
  \let\gplgaddtomacro\g@addto@macro
  % define empty templates for all commands taking text:
  \gdef\gplbacktext{}%
  \gdef\gplfronttext{}%
  \makeatother
  \ifGPblacktext
    % no textcolor at all
    \def\colorrgb#1{}%
    \def\colorgray#1{}%
  \else
    % gray or color?
    \ifGPcolor
      \def\colorrgb#1{\color[rgb]{#1}}%
      \def\colorgray#1{\color[gray]{#1}}%
      \expandafter\def\csname LTw\endcsname{\color{white}}%
      \expandafter\def\csname LTb\endcsname{\color{black}}%
      \expandafter\def\csname LTa\endcsname{\color{black}}%
      \expandafter\def\csname LT0\endcsname{\color[rgb]{1,0,0}}%
      \expandafter\def\csname LT1\endcsname{\color[rgb]{0,1,0}}%
      \expandafter\def\csname LT2\endcsname{\color[rgb]{0,0,1}}%
      \expandafter\def\csname LT3\endcsname{\color[rgb]{1,0,1}}%
      \expandafter\def\csname LT4\endcsname{\color[rgb]{0,1,1}}%
      \expandafter\def\csname LT5\endcsname{\color[rgb]{1,1,0}}%
      \expandafter\def\csname LT6\endcsname{\color[rgb]{0,0,0}}%
      \expandafter\def\csname LT7\endcsname{\color[rgb]{1,0.3,0}}%
      \expandafter\def\csname LT8\endcsname{\color[rgb]{0.5,0.5,0.5}}%
    \else
      % gray
      \def\colorrgb#1{\color{black}}%
      \def\colorgray#1{\color[gray]{#1}}%
      \expandafter\def\csname LTw\endcsname{\color{white}}%
      \expandafter\def\csname LTb\endcsname{\color{black}}%
      \expandafter\def\csname LTa\endcsname{\color{black}}%
      \expandafter\def\csname LT0\endcsname{\color{black}}%
      \expandafter\def\csname LT1\endcsname{\color{black}}%
      \expandafter\def\csname LT2\endcsname{\color{black}}%
      \expandafter\def\csname LT3\endcsname{\color{black}}%
      \expandafter\def\csname LT4\endcsname{\color{black}}%
      \expandafter\def\csname LT5\endcsname{\color{black}}%
      \expandafter\def\csname LT6\endcsname{\color{black}}%
      \expandafter\def\csname LT7\endcsname{\color{black}}%
      \expandafter\def\csname LT8\endcsname{\color{black}}%
    \fi
  \fi
  \setlength{\unitlength}{0.0500bp}%
  \begin{picture}(5040.00,4320.00)%
    \gplgaddtomacro\gplbacktext{%
      \csname LTb\endcsname%
      \put(550,704){\makebox(0,0)[r]{\strut{}$0$}}%
      \put(550,1123){\makebox(0,0)[r]{\strut{}$1$}}%
      \put(550,1542){\makebox(0,0)[r]{\strut{}$2$}}%
      \put(550,1961){\makebox(0,0)[r]{\strut{}$3$}}%
      \put(550,2380){\makebox(0,0)[r]{\strut{}$4$}}%
      \put(550,2798){\makebox(0,0)[r]{\strut{}$5$}}%
      \put(550,3217){\makebox(0,0)[r]{\strut{}$6$}}%
      \put(550,3636){\makebox(0,0)[r]{\strut{}$7$}}%
      \put(550,4055){\makebox(0,0)[r]{\strut{}$8$}}%
      \put(682,484){\makebox(0,0){\strut{}$0$}}%
      \put(1342,484){\makebox(0,0){\strut{}$100$}}%
      \put(2002,484){\makebox(0,0){\strut{}$200$}}%
      \put(2663,484){\makebox(0,0){\strut{}$300$}}%
      \put(3323,484){\makebox(0,0){\strut{}$400$}}%
      \put(3983,484){\makebox(0,0){\strut{}$500$}}%
      \put(4643,484){\makebox(0,0){\strut{}$600$}}%
      \put(176,2379){\rotatebox{-270}{\makebox(0,0){\strut{}$\Delta \Phi$}}}%
      \put(2662,154){\makebox(0,0){\strut{}number of particles, $N$}}%
    }%
    \gplgaddtomacro\gplfronttext{%
      \csname LTb\endcsname%
      \put(2002,3882){\makebox(0,0)[r]{\strut{}$z$ = 20.70}}%
      \csname LTb\endcsname%
      \put(2002,3662){\makebox(0,0)[r]{\strut{}$z$ = 20.75}}%
      \csname LTb\endcsname%
      \put(2002,3442){\makebox(0,0)[r]{\strut{}$z$ = 20.80}}%
      \csname LTb\endcsname%
      \put(2002,3222){\makebox(0,0)[r]{\strut{}$z$ = 20.85}}%
      \csname LTb\endcsname%
      \put(2002,3002){\makebox(0,0)[r]{\strut{}$z$ = 20.90}}%
    }%
    \gplbacktext
    \put(0,0){\includegraphics{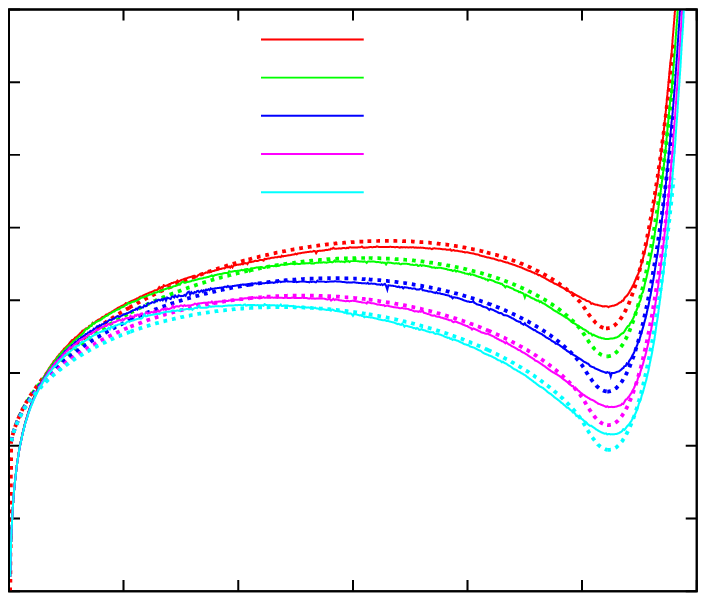}}%
    \gplfronttext
  \end{picture}%
\endgroup

%% file: ld_r_sweep.tex
% GNUPLOT: LaTeX picture with Postscript
\begingroup
  \makeatletter
  \providecommand\color[2][]{%
    \GenericError{(gnuplot) \space\space\space\@spaces}{%
      Package color not loaded in conjunction with
      terminal option `colourtext'%
    }{See the gnuplot documentation for explanation.%
    }{Either use 'blacktext' in gnuplot or load the package
      color.sty in LaTeX.}%
    \renewcommand\color[2][]{}%
  }%
  \providecommand\includegraphics[2][]{%
    \GenericError{(gnuplot) \space\space\space\@spaces}{%
      Package graphicx or graphics not loaded%
    }{See the gnuplot documentation for explanation.%
    }{The gnuplot epslatex terminal needs graphicx.sty or graphics.sty.}%
    \renewcommand\includegraphics[2][]{}%
  }%
  \providecommand\rotatebox[2]{#2}%
  \@ifundefined{ifGPcolor}{%
    \newif\ifGPcolor
    \GPcolortrue
  }{}%
  \@ifundefined{ifGPblacktext}{%
    \newif\ifGPblacktext
    \GPblacktextfalse
  }{}%
  % define a \g@addto@macro without @ in the name:
  \let\gplgaddtomacro\g@addto@macro
  % define empty templates for all commands taking text:
  \gdef\gplbacktext{}%
  \gdef\gplfronttext{}%
  \makeatother
  \ifGPblacktext
    % no textcolor at all
    \def\colorrgb#1{}%
    \def\colorgray#1{}%
  \else
    % gray or color?
    \ifGPcolor
      \def\colorrgb#1{\color[rgb]{#1}}%
      \def\colorgray#1{\color[gray]{#1}}%
      \expandafter\def\csname LTw\endcsname{\color{white}}%
      \expandafter\def\csname LTb\endcsname{\color{black}}%
      \expandafter\def\csname LTa\endcsname{\color{black}}%
      \expandafter\def\csname LT0\endcsname{\color[rgb]{1,0,0}}%
      \expandafter\def\csname LT1\endcsname{\color[rgb]{0,1,0}}%
      \expandafter\def\csname LT2\endcsname{\color[rgb]{0,0,1}}%
      \expandafter\def\csname LT3\endcsname{\color[rgb]{1,0,1}}%
      \expandafter\def\csname LT4\endcsname{\color[rgb]{0,1,1}}%
      \expandafter\def\csname LT5\endcsname{\color[rgb]{1,1,0}}%
      \expandafter\def\csname LT6\endcsname{\color[rgb]{0,0,0}}%
      \expandafter\def\csname LT7\endcsname{\color[rgb]{1,0.3,0}}%
      \expandafter\def\csname LT8\endcsname{\color[rgb]{0.5,0.5,0.5}}%
    \else
      % gray
      \def\colorrgb#1{\color{black}}%
      \def\colorgray#1{\color[gray]{#1}}%
      \expandafter\def\csname LTw\endcsname{\color{white}}%
      \expandafter\def\csname LTb\endcsname{\color{black}}%
      \expandafter\def\csname LTa\endcsname{\color{black}}%
      \expandafter\def\csname LT0\endcsname{\color{black}}%
      \expandafter\def\csname LT1\endcsname{\color{black}}%
      \expandafter\def\csname LT2\endcsname{\color{black}}%
      \expandafter\def\csname LT3\endcsname{\color{black}}%
      \expandafter\def\csname LT4\endcsname{\color{black}}%
      \expandafter\def\csname LT5\endcsname{\color{black}}%
      \expandafter\def\csname LT6\endcsname{\color{black}}%
      \expandafter\def\csname LT7\endcsname{\color{black}}%
      \expandafter\def\csname LT8\endcsname{\color{black}}%
    \fi
  \fi
  \setlength{\unitlength}{0.0500bp}%
  \begin{picture}(5040.00,4320.00)%
    \gplgaddtomacro\gplbacktext{%
      \csname LTb\endcsname%
      \put(550,704){\makebox(0,0)[r]{\strut{}$0$}}%
      \put(550,1123){\makebox(0,0)[r]{\strut{}$1$}}%
      \put(550,1542){\makebox(0,0)[r]{\strut{}$2$}}%
      \put(550,1961){\makebox(0,0)[r]{\strut{}$3$}}%
      \put(550,2380){\makebox(0,0)[r]{\strut{}$4$}}%
      \put(550,2798){\makebox(0,0)[r]{\strut{}$5$}}%
      \put(550,3217){\makebox(0,0)[r]{\strut{}$6$}}%
      \put(550,3636){\makebox(0,0)[r]{\strut{}$7$}}%
      \put(550,4055){\makebox(0,0)[r]{\strut{}$8$}}%
      \put(682,484){\makebox(0,0){\strut{}$0$}}%
      \put(1210,484){\makebox(0,0){\strut{}$200$}}%
      \put(1738,484){\makebox(0,0){\strut{}$400$}}%
      \put(2266,484){\makebox(0,0){\strut{}$600$}}%
      \put(2795,484){\makebox(0,0){\strut{}$800$}}%
      \put(3323,484){\makebox(0,0){\strut{}$1000$}}%
      \put(3851,484){\makebox(0,0){\strut{}$1200$}}%
      \put(4379,484){\makebox(0,0){\strut{}$1400$}}%
      \put(176,2379){\rotatebox{-270}{\makebox(0,0){\strut{}$\Delta \Phi$}}}%
      \put(2662,154){\makebox(0,0){\strut{}number of particles, $N$}}%
    }%
    \gplgaddtomacro\gplfronttext{%
      \csname LTb\endcsname%
      \put(3656,3882){\makebox(0,0)[r]{\strut{}$R$ = 4$r_{m}$}}%
      \csname LTb\endcsname%
      \put(3656,3662){\makebox(0,0)[r]{\strut{}$R$ = 5$r_{m}$}}%
      \csname LTb\endcsname%
      \put(3656,3442){\makebox(0,0)[r]{\strut{}$R$ = 6$r_{m}$}}%
      \csname LTb\endcsname%
      \put(3656,3222){\makebox(0,0)[r]{\strut{}$R$ = 7$r_{m}$}}%
      \csname LTb\endcsname%
      \put(3656,3002){\makebox(0,0)[r]{\strut{}$R$ = 8$r_{m}$}}%
      \csname LTb\endcsname%
      \put(3656,2782){\makebox(0,0)[r]{\strut{}$R$ = 9$r_{m}$}}%
      \csname LTb\endcsname%
      \put(3656,2562){\makebox(0,0)[r]{\strut{}$R$ = 10$r_{m}$}}%
    }%
    \gplbacktext
    \put(0,0){\includegraphics{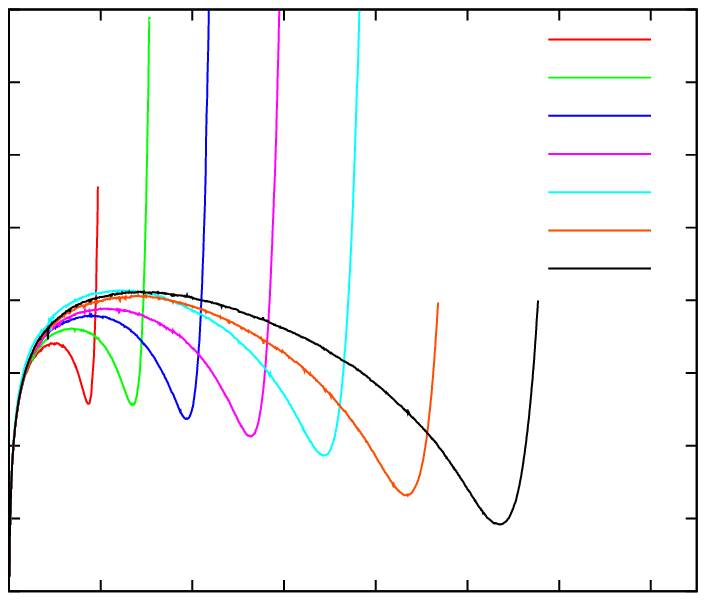}}%
    \gplfronttext
  \end{picture}%
\endgroup

%% file: crit_radius.tex
% GNUPLOT: LaTeX picture with Postscript
\begingroup
  \makeatletter
  \providecommand\color[2][]{%
    \GenericError{(gnuplot) \space\space\space\@spaces}{%
      Package color not loaded in conjunction with
      terminal option `colourtext'%
    }{See the gnuplot documentation for explanation.%
    }{Either use 'blacktext' in gnuplot or load the package
      color.sty in LaTeX.}%
    \renewcommand\color[2][]{}%
  }%
  \providecommand\includegraphics[2][]{%
    \GenericError{(gnuplot) \space\space\space\@spaces}{%
      Package graphicx or graphics not loaded%
    }{See the gnuplot documentation for explanation.%
    }{The gnuplot epslatex terminal needs graphicx.sty or graphics.sty.}%
    \renewcommand\includegraphics[2][]{}%
  }%
  \providecommand\rotatebox[2]{#2}%
  \@ifundefined{ifGPcolor}{%
    \newif\ifGPcolor
    \GPcolortrue
  }{}%
  \@ifundefined{ifGPblacktext}{%
    \newif\ifGPblacktext
    \GPblacktextfalse
  }{}%
  % define a \g@addto@macro without @ in the name:
  \let\gplgaddtomacro\g@addto@macro
  % define empty templates for all commands taking text:
  \gdef\gplbacktext{}%
  \gdef\gplfronttext{}%
  \makeatother
  \ifGPblacktext
    % no textcolor at all
    \def\colorrgb#1{}%
    \def\colorgray#1{}%
  \else
    % gray or color?
    \ifGPcolor
      \def\colorrgb#1{\color[rgb]{#1}}%
      \def\colorgray#1{\color[gray]{#1}}%
      \expandafter\def\csname LTw\endcsname{\color{white}}%
      \expandafter\def\csname LTb\endcsname{\color{black}}%
      \expandafter\def\csname LTa\endcsname{\color{black}}%
      \expandafter\def\csname LT0\endcsname{\color[rgb]{1,0,0}}%
      \expandafter\def\csname LT1\endcsname{\color[rgb]{0,1,0}}%
      \expandafter\def\csname LT2\endcsname{\color[rgb]{0,0,1}}%
      \expandafter\def\csname LT3\endcsname{\color[rgb]{1,0,1}}%
      \expandafter\def\csname LT4\endcsname{\color[rgb]{0,1,1}}%
      \expandafter\def\csname LT5\endcsname{\color[rgb]{1,1,0}}%
      \expandafter\def\csname LT6\endcsname{\color[rgb]{0,0,0}}%
      \expandafter\def\csname LT7\endcsname{\color[rgb]{1,0.3,0}}%
      \expandafter\def\csname LT8\endcsname{\color[rgb]{0.5,0.5,0.5}}%
    \else
      % gray
      \def\colorrgb#1{\color{black}}%
      \def\colorgray#1{\color[gray]{#1}}%
      \expandafter\def\csname LTw\endcsname{\color{white}}%
      \expandafter\def\csname LTb\endcsname{\color{black}}%
      \expandafter\def\csname LTa\endcsname{\color{black}}%
      \expandafter\def\csname LT0\endcsname{\color{black}}%
      \expandafter\def\csname LT1\endcsname{\color{black}}%
      \expandafter\def\csname LT2\endcsname{\color{black}}%
      \expandafter\def\csname LT3\endcsname{\color{black}}%
      \expandafter\def\csname LT4\endcsname{\color{black}}%
      \expandafter\def\csname LT5\endcsname{\color{black}}%
      \expandafter\def\csname LT6\endcsname{\color{black}}%
      \expandafter\def\csname LT7\endcsname{\color{black}}%
      \expandafter\def\csname LT8\endcsname{\color{black}}%
    \fi
  \fi
  \setlength{\unitlength}{0.0500bp}%
  \begin{picture}(5040.00,4320.00)%
    \gplgaddtomacro\gplbacktext{%
      \csname LTb\endcsname%
      \put(814,704){\makebox(0,0)[r]{\strut{}$0$}}%
      \put(814,1076){\makebox(0,0)[r]{\strut{}$0.2$}}%
      \put(814,1449){\makebox(0,0)[r]{\strut{}$0.4$}}%
      \put(814,1821){\makebox(0,0)[r]{\strut{}$0.6$}}%
      \put(814,2193){\makebox(0,0)[r]{\strut{}$0.8$}}%
      \put(814,2566){\makebox(0,0)[r]{\strut{}$1$}}%
      \put(814,2938){\makebox(0,0)[r]{\strut{}$1.2$}}%
      \put(814,3310){\makebox(0,0)[r]{\strut{}$1.4$}}%
      \put(814,3683){\makebox(0,0)[r]{\strut{}$1.6$}}%
      \put(814,4055){\makebox(0,0)[r]{\strut{}$1.8$}}%
      \put(946,484){\makebox(0,0){\strut{}$0$}}%
      \put(1357,484){\makebox(0,0){\strut{}$0.2$}}%
      \put(1768,484){\makebox(0,0){\strut{}$0.4$}}%
      \put(2178,484){\makebox(0,0){\strut{}$0.6$}}%
      \put(2589,484){\makebox(0,0){\strut{}$0.8$}}%
      \put(3000,484){\makebox(0,0){\strut{}$1$}}%
      \put(3411,484){\makebox(0,0){\strut{}$1.2$}}%
      \put(3821,484){\makebox(0,0){\strut{}$1.4$}}%
      \put(4232,484){\makebox(0,0){\strut{}$1.6$}}%
      \put(4643,484){\makebox(0,0){\strut{}$1.8$}}%
      \put(176,2379){\rotatebox{-270}{\makebox(0,0){\strut{}$r_c/R$}}}%
      \put(2794,154){\makebox(0,0){\strut{}$\arctan{\left(r^0_{\rm c}/R\right)}$}}%
    }%
    \gplgaddtomacro\gplfronttext{%
      \csname LTb\endcsname%
      \put(3656,1317){\makebox(0,0)[r]{\strut{}$T$ = 0.391, $z_{\rm ref}$ = 19.0}}%
      \csname LTb\endcsname%
      \put(3656,1097){\makebox(0,0)[r]{\strut{}$T$ = 0.396, $z_{\rm ref}$ = 20.6}}%
      \csname LTb\endcsname%
      \put(3656,877){\makebox(0,0)[r]{\strut{}$T$ =  0.400, $z_{\rm ref}$ = 20.9}}%
    }%
    \gplbacktext
    \put(0,0){\includegraphics{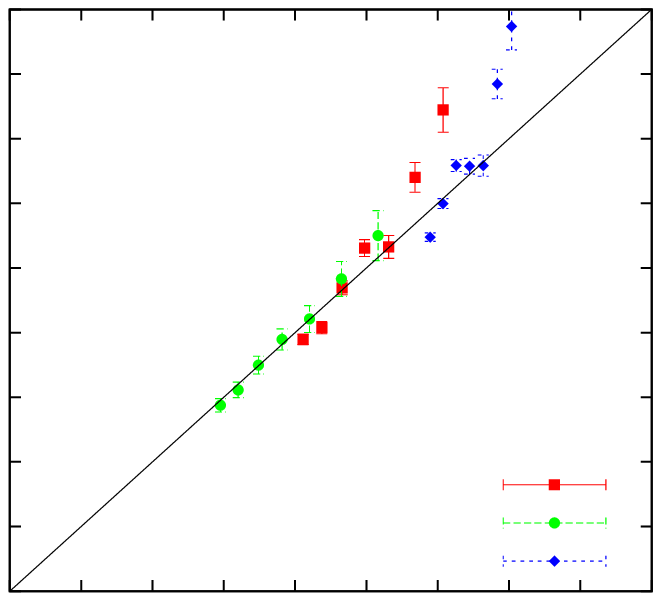}}%
    \gplfronttext
  \end{picture}%
\endgroup

%% file: ld_canon.tex
% GNUPLOT: LaTeX picture with Postscript
\begingroup
  \makeatletter
  \providecommand\color[2][]{%
    \GenericError{(gnuplot) \space\space\space\@spaces}{%
      Package color not loaded in conjunction with
      terminal option `colourtext'%
    }{See the gnuplot documentation for explanation.%
    }{Either use 'blacktext' in gnuplot or load the package
      color.sty in LaTeX.}%
    \renewcommand\color[2][]{}%
  }%
  \providecommand\includegraphics[2][]{%
    \GenericError{(gnuplot) \space\space\space\@spaces}{%
      Package graphicx or graphics not loaded%
    }{See the gnuplot documentation for explanation.%
    }{The gnuplot epslatex terminal needs graphicx.sty or graphics.sty.}%
    \renewcommand\includegraphics[2][]{}%
  }%
  \providecommand\rotatebox[2]{#2}%
  \@ifundefined{ifGPcolor}{%
    \newif\ifGPcolor
    \GPcolortrue
  }{}%
  \@ifundefined{ifGPblacktext}{%
    \newif\ifGPblacktext
    \GPblacktextfalse
  }{}%
  % define a \g@addto@macro without @ in the name:
  \let\gplgaddtomacro\g@addto@macro
  % define empty templates for all commands taking text:
  \gdef\gplbacktext{}%
  \gdef\gplfronttext{}%
  \makeatother
  \ifGPblacktext
    % no textcolor at all
    \def\colorrgb#1{}%
    \def\colorgray#1{}%
  \else
    % gray or color?
    \ifGPcolor
      \def\colorrgb#1{\color[rgb]{#1}}%
      \def\colorgray#1{\color[gray]{#1}}%
      \expandafter\def\csname LTw\endcsname{\color{white}}%
      \expandafter\def\csname LTb\endcsname{\color{black}}%
      \expandafter\def\csname LTa\endcsname{\color{black}}%
      \expandafter\def\csname LT0\endcsname{\color[rgb]{1,0,0}}%
      \expandafter\def\csname LT1\endcsname{\color[rgb]{0,1,0}}%
      \expandafter\def\csname LT2\endcsname{\color[rgb]{0,0,1}}%
      \expandafter\def\csname LT3\endcsname{\color[rgb]{1,0,1}}%
      \expandafter\def\csname LT4\endcsname{\color[rgb]{0,1,1}}%
      \expandafter\def\csname LT5\endcsname{\color[rgb]{1,1,0}}%
      \expandafter\def\csname LT6\endcsname{\color[rgb]{0,0,0}}%
      \expandafter\def\csname LT7\endcsname{\color[rgb]{1,0.3,0}}%
      \expandafter\def\csname LT8\endcsname{\color[rgb]{0.5,0.5,0.5}}%
    \else
      % gray
      \def\colorrgb#1{\color{black}}%
      \def\colorgray#1{\color[gray]{#1}}%
      \expandafter\def\csname LTw\endcsname{\color{white}}%
      \expandafter\def\csname LTb\endcsname{\color{black}}%
      \expandafter\def\csname LTa\endcsname{\color{black}}%
      \expandafter\def\csname LT0\endcsname{\color{black}}%
      \expandafter\def\csname LT1\endcsname{\color{black}}%
      \expandafter\def\csname LT2\endcsname{\color{black}}%
      \expandafter\def\csname LT3\endcsname{\color{black}}%
      \expandafter\def\csname LT4\endcsname{\color{black}}%
      \expandafter\def\csname LT5\endcsname{\color{black}}%
      \expandafter\def\csname LT6\endcsname{\color{black}}%
      \expandafter\def\csname LT7\endcsname{\color{black}}%
      \expandafter\def\csname LT8\endcsname{\color{black}}%
    \fi
  \fi
  \setlength{\unitlength}{0.0500bp}%
  \begin{picture}(5040.00,4320.00)%
    \gplgaddtomacro\gplbacktext{%
      \csname LTb\endcsname%
      \put(814,704){\makebox(0,0)[r]{\strut{}$-2$}}%
      \put(814,1374){\makebox(0,0)[r]{\strut{}$0$}}%
      \put(814,2044){\makebox(0,0)[r]{\strut{}$2$}}%
      \put(814,2715){\makebox(0,0)[r]{\strut{}$4$}}%
      \put(814,3385){\makebox(0,0)[r]{\strut{}$6$}}%
      \put(814,4055){\makebox(0,0)[r]{\strut{}$8$}}%
      \put(946,484){\makebox(0,0){\strut{}$0$}}%
      \put(1657,484){\makebox(0,0){\strut{}$50$}}%
      \put(2368,484){\makebox(0,0){\strut{}$100$}}%
      \put(3079,484){\makebox(0,0){\strut{}$150$}}%
      \put(3790,484){\makebox(0,0){\strut{}$200$}}%
      \put(4501,484){\makebox(0,0){\strut{}$250$}}%
      \put(176,2379){\rotatebox{-270}{\makebox(0,0){\strut{}$\Delta F$}}}%
      \put(2794,154){\makebox(0,0){\strut{}number of particles, $N$}}%
    }%
    \gplgaddtomacro\gplfronttext{%
      \csname LTb\endcsname%
      \put(2134,1537){\makebox(0,0)[r]{\strut{}$N_0$ = 100}}%
      \csname LTb\endcsname%
      \put(2134,1317){\makebox(0,0)[r]{\strut{}$N_0$ = 150}}%
      \csname LTb\endcsname%
      \put(2134,1097){\makebox(0,0)[r]{\strut{}$N_0$ = 200}}%
      \csname LTb\endcsname%
      \put(2134,877){\makebox(0,0)[r]{\strut{}$N_0$ = 250}}%
    }%
    \gplbacktext
    \put(0,0){\includegraphics{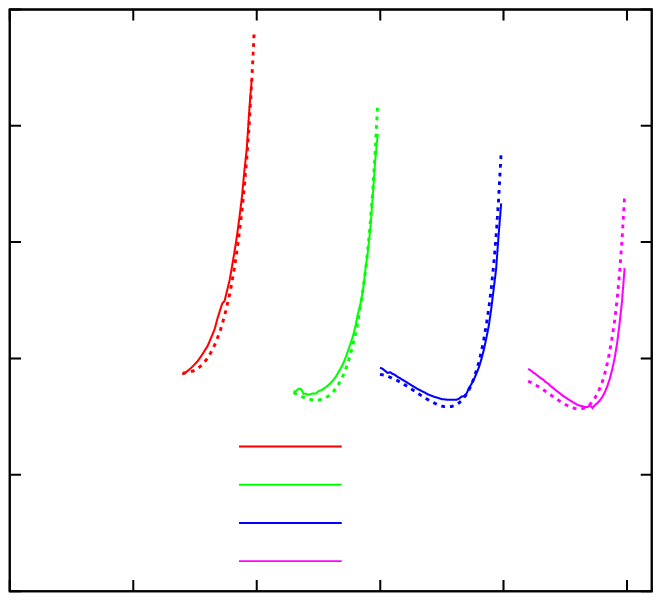}}%
    \gplfronttext
  \end{picture}%
\endgroup